\documentclass[onecolumn, draft, 12pt]{IEEEtran}
\usepackage{amsmath,cases}
\usepackage{amsthm}
\usepackage{amssymb}
\usepackage{cite}
\usepackage{amsfonts}
\usepackage{enumerate}
\usepackage[final]{graphicx}

\newcommand{\ttheta}{\ensuremath{\theta}}
\newcommand{\isrx}{\ensuremath{x_i}}
\newcommand{\ieta}{\ensuremath{\eta_i}}
\newcommand{\thetarng}{\ensuremath{\theta_R}}

\newcommand{\frx}{\ensuremath{y_{{L}}}}
\newcommand{\feta}{\ensuremath{v}}
\newcommand{\anffac}{\ensuremath{\alpha}}

\newcommand{\powcnst}{\ensuremath{P_T}}
\newcommand{\nrmfrx}{\ensuremath{z_L}}
\newcommand{\nrmfrxI}{\ensuremath{z_L^I}}
\newcommand{\nrmfrxR}{\ensuremath{z_L^R}}
\newcommand{\nrmfeta}{\ensuremath{\frac{v}{\sqrt{L}}}}
\newcommand{\esttheta}{\ensuremath{\widehat{\theta}}}
\newcommand{\Chareta}{\ensuremath{\varphi_{\eta}(\omega)}}
\newcommand{\Chardeta}{\ensuremath{\varphi_{\eta}(2\omega)}}
\newcommand{\Charseta}{\ensuremath{\varphi_{\eta}^2(\omega)}}
\newcommand{\avgnrmfrx}{\ensuremath{\overline{z}}}
\newcommand{\kurt}{\ensuremath{\kappa_{\eta}}}
\newcommand{\betasol}{\ensuremath{\beta^{\ast}_{\rm G}}}
\newcommand{\sigeta}{\ensuremath{\sigma_{\eta}^2}}
\newcommand{\sigv}{\ensuremath{\sigma_v^2}}
\newcommand{\sigx}{\ensuremath{\sigma_X^2}}
\newcommand{\w}{\ensuremath{\omega}}
\newcommand{\asv}{\ensuremath{AsV(\w)}}
\newcommand{\bt}{\ensuremath{\beta}}
\newcommand{\gm}{\ensuremath{\gamma}}

\newtheorem{thm}{Theorem}

\begin{document}

\title{Robust Distributed Estimation over Multiple Access Channels with Constant Modulus Signaling}
\author{Cihan Tepedelenlio\u{g}lu, \emph{Member, IEEE}, Adarsh B. Narasimhamurthy 
\thanks{The authors are with the School of Electrical, Computer, and Energy Engineering, Arizona
State University, Tempe, AZ 85287, USA. (Email:
\{cihan,anbangal\}@asu.edu).} } \maketitle

\begin{abstract}
A distributed estimation scheme where the sensors transmit with constant modulus signals
over a multiple access channel is considered. The proposed estimator is shown to be
strongly consistent for any sensing noise distribution in the i.i.d. case both for a per-sensor
power constraint, and a total power constraint.
When the distributions of the sensing noise are not identical, a bound on the variances
is shown to establish strong consistency. The estimator is shown to be asymptotically
normal with a variance (AsV) that depends on the characteristic function of the sensing noise.
Optimization of the AsV is considered with respect to a transmission phase parameter for
a variety of noise distributions exhibiting differing levels of impulsive behavior.
The robustness of the estimator to impulsive sensing noise distributions such as those with
positive excess kurtosis, or those that do not have finite moments is shown. The proposed
estimator is favorably compared with the amplify and forward scheme under an impulsive noise
scenario. The effect of fading is
shown to not affect the consistency of the estimator, but to scale the
asymptotic variance by a constant fading penalty depending on the
fading statistics. Simulations corroborate our analytical results.
\end{abstract}


\section{Introduction} \label{sec:intro}
In inference-based wireless sensor networks, low-power sensors with limited battery
and peak-power capabilities transmit their observations to a fusion center (FC) for
detection of events or estimation of parameters. For distributed estimation, much of the
literature has focused on a set of orthogonal (parallel) fading channels between
the sensors and the FC (please see \cite{magazine} and the references therein).
The bandwidth requirements of such an orthogonal
WSN scales linearly with the number of sensors. In contrast, 
over multiple access channels where the sensor transmissions are
simultaneous and in the same frequency band, the utilized bandwidth
does not depend on the number of sensors. In both cases, sensors may
adopt either a digital or analog method for relaying the sensed
information to the FC. The digital method consists of quantizing the
sensed data and transmitting with digital modulation over a
rate-constrained channel. In these cases, the required channel
bandwidth is proportional number of bits at the output of the
quantizer which are transmitted after pulse shaping and digital
modulation. The analog method consists of transmitting unquantized
data by appropriately pulse shaping and amplitude or phase
modulating to consume finite bandwidth.

The literature on distributed estimation over multiple access channels
has mainly involved analog sensor transmission schemes 
where the instantaneous transmit power is influenced by the sensor measurement noise and
is not bounded \cite{gastpar2,cui,maheshicassp,maheshasilomar,jaya,mergen,willet}.
In \cite{gastpar2}, distributed estimation over Gaussian multiple access channels
is studied from a joint source-channel coding point of view.
Reference \cite{cui} considers optimization of the sensor gains in
the presence of channel fading. In \cite{maheshicassp} and
\cite{maheshasilomar}, the effects of different fading distributions
and channel feedback on the performance of distributed estimators
over multiple access channels is studied. A direct-sequence CDMA
with amplify and forward (AF) is considered in \cite{jaya}, where
the asymptotic MSE is studied. In \cite{mergen}, the authors
introduce a type-based multiple access scheme where more than one
orthogonal channel is utilized albeit less in number than the number
of sensors. In \cite{willet}, a likelihood-based multiple access
approach is introduced. The latter two references do not explicitly
estimate a location parameter (such as the mean or the median) of
the sensed data. In these aforementioned schemes, the sensor power
management issues arising from the dependence of the instantaneous
transmit power on the sensing noise have not been addressed.
Moreover, for sensors operating in adverse conditions, robustness to
impulsive  noise \footnote{referring to distributions whose tails
decay slower than that of Gaussian noise} is of paramount
importance, which has not been addressed in the literature in the
context of distributed estimation over multiple access channels.

In this work, a distributed estimation scheme is considered 
where the sensor transmissions have constant modulus with fixed
instantaneous transmit power. The proposed estimator is universal in
the sense of \cite{universal0} (or ``distribution-free'' in
statistical parlance) in that the estimator does not depend on the
distribution or the parameters of the sensing or channel noise.
Unlike the orthogonal framework in \cite{universal0}, multiple
access channels are considered herein, and the sensing noise is not
assumed bounded. The estimator is shown to be strongly consistent
for any noise distribution, including those with no finite moments,
in the i.i.d. case. The distribution-free aspect is also very useful
in heterogenious scenarios where several different kinds of noise
are simultaneously present, such as additive Gaussian noise along
with quantization noise.

The sensors transmit with constant modulus transmissions whose phase
is linear with the sensed data. The FC estimates a common location
parameter (such as the mean, or the median) of the sensed signal
where the sensing noise samples are not assumed to be identically
distributed, or from any specific distribution. It is shown that the
proposed estimator is strongly consistent even when the sensing
noise is not identically distributed, provided that their variances
are bounded. While the estimator is shown to be consistent in this
general framework, the asymptotic variance of the estimator is
derived for the i.i.d. sensing noise and shown to depend on its
characteristic function (CF). Upper bounds on, and optimization of
the asymptotic variance with the transmit phase parameter $\w$ is
considered for different distributions on the sensing noise
including impulsive ones. The proposed estimator is compared with
AF, where the robustness of the proposed estimator is highlighted.
The effect of fading is shown to not affect the consistency of the
estimator, but only to scale the asymptotic variance by a constant
fading penalty depending on the fading statistics.

\section{System Model} \label{sec: model}
Consider the sensing model, with $L$ sensors,
\begin{equation}\label{eqn: ith_sensor_rx}
\isrx = \ttheta+\ieta  \hspace{1 in} i = 1, \ldots, L
\end{equation}where $\ttheta$ is an unknown real-valued parameter in a bounded interval $[0,\hspace{0.05 in} \thetarng]$ of known length, $\thetarng<\infty$, $\ieta$ are a
mutually independent, symmetric real-valued noise with zero median (i.e., its pdf, when it exists,  is symmetric about
zero), and $\isrx$ is the measurement at the $i^{th}$ sensor. Note that $\ieta$
are not necessarily identically distributed, bounded, and need not have finite moments.
We consider a setting where the $i^{th}$ sensor transmits its measurement using a constant modulus signal
$\sqrt{\rho}e^{j\w\isrx}$ over a Gaussian multiple access channel so that the received signal at the fusion center (FC) is given by

\begin{equation}\label{eqn: fusion_cntr_rx}
\frx = \sqrt{\rho} \sum_{i=1}^{L} e^{j\w \isrx}+\feta
\end{equation}where the transmitted signal at each sensor has a per-sensor power
of $\rho$, $0<\w \leq 2\pi/\thetarng$ is a design parameter to be
optimized, and $\feta$ is additive noise. Note that the restriction
$\w \in (0, 2\pi/\thetarng]$ is necessary even in the absence of
sensing and channel noise, to uniquely determine $\ttheta$ from
$\frx$. Estimation in a single time snap shot is considered, which
is why the time index is dropped. The transmitted signal has a
deterministic fixed power $\rho$ which does not suffer from the
problems of random transmit power seen in AF schemes where the
transmitted signal from the $i^{th}$ sensor is
given by $\anffac \isrx = \anffac(\ttheta+\ieta)$ with instantaneous power per sensor $\alpha^2 (\ttheta+\ieta)^2$, which is an unbounded random variable (RV) when $\ieta$ is. In AF transmission, $\alpha$ is a coefficient which might depend on the sensor index, as well as
on $L$ through a power constraint, but does not depend on $\isrx$ \cite{cui2007,gastparb2003}.
Note that the total transmit power from all the sensors in (\ref{eqn: fusion_cntr_rx}) is $\rho L$. We begin by considering a fixed total power constraint $\powcnst$ implying that the per-sensor power $\rho{=}\powcnst/L$ is a function of $L$. Later, in Section \ref{sec: per_sensor_pow_cnst}, we will also consider a fixed per-sensor power scheme where $\rho$ will not be a function of the number of sensors $L$, which implies $\powcnst \rightarrow \infty$ as $L \rightarrow \infty$.

\section{The Estimation Problem} \label{sec: Estimatino_problem}

We would like to estimate $\ttheta$ from $\frx$ which under the
total power constraint is given by

\begin{equation}
\frx = e^{j \w \ttheta} \sqrt{\frac{\powcnst}{L}}\sum_{i=1}^{L}e^{j \w \ieta}+\feta .
\end{equation}
We do not assume that $\ieta$ are identically distributed, or that $\ieta$ are from any
specific distribution since
a universal estimator which is independent of the distribution of $\ieta$ is desired. Let,

\begin{equation} \label{emp}
\nrmfrx := \frac{y_L}{\sqrt{L}} = e^{j \w \ttheta} \sqrt{\powcnst} \frac{1}{L} \sum_{i=1}^{L} e^{j \w \ieta}+ \nrmfeta \;,
\end{equation}
and define $\varphi_{\ieta}(\w):={\rm{E}}\left[e^{j \eta_i
\w}\right]$ as the CF of $\ieta$. Due to the law of large numbers we
have
\begin{equation} \label{phi}
\frac{1}{L} \sum_{i=1}^L e^{j \eta_i \w} \rightarrow \varphi(\w):= \lim_{L \rightarrow \infty}
\frac{1}{L} \sum_{i=1}^L \varphi_{\ieta}(\w)
\end{equation}
(where $\rightarrow$ indicates convergence almost surely), and we
use the fact that the variances ${\rm var}(e^{j \eta_i \w}) = 1-
\varphi^2_{\ieta}(\w) \leq 1$ are bounded to invoke Kolmogorov's
strong law of large numbers for non-identically distributed RVs
\cite[pp. 259]{feller}. Since $\ieta$ are symmetric,
$\{\varphi_{\ieta}(\w)\}$ are real-valued and therefore
$\varphi(\w)$ is also real-valued.

Consider the conditions under which $\varphi(\w)$ is a CF, which
will be important in the consistency of the proposed estimator.
Since convex combinations of CFs are CFs \cite{ushakov1999}, the
partial sums $L^{-1} \sum_{i=1}^L \varphi_{\ieta}(\w)$ are as well.
From the continuity theorem \cite[Corollary 1.2.2]{ushakov1999} if a
sequence of CFs converges pointwise to a function continuous at
$\w=0$, then the limit is a CF. Therefore $\varphi(\w)$ in
(\ref{phi}) is a CF if $\varphi(\w)$ is continuous at $\w=0$.

The natural estimator that we will adopt is based on the phase of $\nrmfrx$:
\begin{equation} \label{eqn: prop_estimator}
\esttheta = \frac{1}{\w} \tan^{-1}\left(\frac{\nrmfrxI}{\nrmfrxR} \right) ,
\end{equation} where $\nrmfrxR {:=} {\rm{Re}}\{\nrmfrx\}$ and $\nrmfrxI {:=} {\rm{Im}}\{\nrmfrx \}$.
Note that this estimator does not depend on the distributions of $\ieta$ or $\feta$, as desired.
We now establish the strong consistency of the proposed estimator $\esttheta$:

\begin{thm} \label{thm1}
The estimator $\esttheta$ in (\ref{eqn: prop_estimator}) is strongly consistent provided that $\w \in (0,2\pi/\theta_R]$ is chosen to satisfy $\varphi(\w) \neq 0$.
\end{thm}

\begin{IEEEproof}
Taking the real and imaginary parts of (\ref{emp}) and (\ref{phi}) due to the strong law of large numbers
$\nrmfrxR \rightarrow \avgnrmfrx^R := \sqrt{\powcnst}\cos (\w \ttheta) \varphi(\w)$ and $\nrmfrxI \rightarrow \avgnrmfrx^I := \sqrt{\powcnst}\sin(\w \ttheta) \varphi(\w)$ almost surely. Since $\esttheta$ in (\ref{eqn: prop_estimator}) is a continuous function of $\left[\nrmfrxR \hspace{0.1 in} \nrmfrxI \right]$, $\esttheta \rightarrow ({1/\w}) \tan^{-1} \left( {\avgnrmfrx^I}/{\avgnrmfrx^R} \right)=\ttheta$ almost surely \cite[Thm 3.14]{porat1994}. We need the assumption
that $\varphi(\w) \neq 0$ since otherwise $\ttheta$ cannot be uniquely determined from $\avgnrmfrx^R$ and $\avgnrmfrx^I$.
\end{IEEEproof}

We now investigate when an $\w$ that satisfies the conditions of Theorem \ref{thm1} exists. Consider
first the identically distributed case where $\ieta$ have a common distribution with a RV $\eta$ so that
$\varphi(\w) = \varphi_{\eta}(\w)$ is a CF.
Many distributions such as Gaussian, Laplace, and Cauchy satisfy $\varphi_{\eta}(\w)>0$ for all $\w$.
If the common sensing noise distribution is known to have this property, then any choice of $\w \in (0,2\pi/\theta_R]$
would clearly satisfy the conditions of Theorem \ref{thm1}. In the more general case, where
nothing is known or assumed about $\eta$, a sufficiently small $\w$ satisfies $\varphi(\w)>0$
since all CFs at the origin are equal to 1 and continuous.
So, for identically distributed sensing noise, an $\w$ 
for which (\ref{eqn: prop_estimator}) is strongly consistent can always be found, even if the sensing noise variance does not exist.

In the general non-identically distributed case, this argument does
not follow since $\varphi(\w)$ is not necessarily a CF. However, if
$\varphi(\w)$ is continuous at $\w=0$, it is a CF by the continuity
theorem \cite{ushakov1999} and the argument above follows. For an
example of when $\varphi(\w)$ is not a CF and not continuous at
$\w=0$, consider a case where $\sum_{i=1}^\infty \varphi_{\ieta}(\w)
< \infty$ for all $\w>0$ such as when $\ieta$ are Gaussian with
variances that depend on $i$ linearly: $\varphi_{\ieta}(\w) =
e^{-\sigma_{i}^2 \w^2/2}$ where $\sigma_{i}^2 = i \sigma^2$, and
$\sum_{i=1}^\infty \varphi_{\ieta}(\w)=(1-\exp(-\sigma^2
\w^2/2))^{-1}< \infty$ by the geometric sum formula.  In this case
due to the $L^{-1}$ factor in (\ref{phi}), $\varphi(\w)=0$ when $\w
>0$, and $\varphi(0)=1$. For this example, $\varphi(\w)$ is not a CF
for any distribution, and there exists no $\w$ that satisfies the
requirements of Theorem \ref{thm1}. Clearly, this is a very severe
case where the sensing noise variance increases linearly with the
sensor index, without bound. In fact, the example above can be
generalized to distributions other than Gaussian, and variances
going to infinity even slower than linearly. For absolutely
continuous sensing noise distributions, when $\ieta$ are expressed
as a scalar multiple of an underlying random variable, and these
scalars (which are proportional to standard deviations when they
exist) go to infinity, it can be shown that the estimator in
(\ref{eqn: prop_estimator}) is not consistent, which is proved next.
\begin{thm} \label{notconsistent}
Let the sensing noise at the $i^{th}$ sensor be a scaled version of a RV $\eta$
with absolutely continuous distribution so that $\eta_i = \sigma_i \eta$ and $\varphi_{\eta_i}(\omega) = \varphi_{\eta}(\sigma_i \omega)$. Suppose also that
$\lim_{i \rightarrow \infty} \sigma_i = \infty$. Then there is
no $\w$ that satisfies the conditions of Theorem \ref{thm1}.
\end{thm}
\begin{IEEEproof}
Recalling from (\ref{phi}) the definition of $\varphi(\w)$, we would
like to show that $\varphi(\w):=\lim_{L \rightarrow \infty} L^{-1}
\sum_{i=1}^L \varphi_{\eta}(\sigma_i \omega) = 0$ for $\w >0$. Since
$\eta$ has an absolutely continuous distribution, $\lim_{x
\rightarrow \infty} \varphi_{\eta}(x)=0$, and because $\lim_{i
\rightarrow \infty} \sigma_i = \infty$, it follows that $\lim_{i \rightarrow \infty}
\varphi_{\eta}(\sigma_i \omega) =0$ for $\w>0$. From \cite[pp.
411]{porat1994} we know that if a sequence satisfies $\lim_{i \rightarrow
\infty} a_i = 0$ then $\lim_{L \rightarrow \infty}
L^{-1}\sum_{i=1}^L a_i=0$, which gives us the proof when applied to
the sequence $\varphi_{\eta}(\sigma_i \omega)$.
\end{IEEEproof}

The following theorem  can loosely be regarded as a converse to Theorem \ref{notconsistent}
and shows that the estimator in (\ref{eqn: prop_estimator}) is consistent when
the variances $\sigma_i^2$ exist and are bounded.
\footnote{It is not a true converse
for two main reasons: (i) $\lim_{i \rightarrow \infty} \sigma_i =
\infty$ required by Theorem \ref{notconsistent} is not the opposite
of $\sigma_i$ being bounded, which is required by Theorem \ref{varbounded},
since it is possible that neither may occur; (ii) Theorem
\ref{notconsistent} requires absolute continuity whereas Theorem
\ref{varbounded} does not.}
\begin{thm} \label{varbounded}
Let ${\rm var}(\ieta)$ exist for all $i$ and $\sigma_{\rm max} :=
\sup_{i} ({\rm var}(\ieta))^{1/2}$ be finite. Then any $0 < \w <
\min(2\pi/\theta_R,\sqrt{2}/\sigma_{\rm max})$ satisfies
$\varphi(\w)>0$, thereby fulfilling the requirement of Theorem
\ref{thm1} on $\w$.
\end{thm}
\begin{IEEEproof}
From \cite[pp. 89]{ushakov1999} we have $\varphi_{\ieta}(\w) \geq 1-
\sigma_{i}^2 \w^2/2$ for any CF with finite variance. Using
(\ref{phi}) we have $\varphi(\w) \geq  1- (\lim_{L \rightarrow
\infty} L^{-1}\sum_{i=1}^L \sigma_{i}^2) \w^2/2 \geq 1- \sigma_{\rm
max}^2 \w^2/2 >0$ where the last inequality holds provided that $\w
< \sqrt{2}/\sigma_{\rm max}$. Since also $\w \leq 2\pi/\theta_R$ we
have the theorem.
\end{IEEEproof}

The estimator in (\ref{eqn: prop_estimator}) relies on constant modulus transmissions from the sensors to the FC, and is strongly consistent over a wide range of scenarios outlined above.
However, the performance of $\esttheta$ will
depend on statistical assumptions on $\{\ieta\}$ and $v$.
The following theorem characterizes this performance, under the assumption that $v \sim \mathcal{C} \mathcal{N} (0, \sigv)$ and
$\{\eta_i\}$ are identically distributed with an arbitrary common distribution.

\begin{thm}\label{thm2}
$\sqrt{L}\left( \esttheta-\ttheta \right)$ is asymptotically normal with zero mean and variance given by,

\begin{equation} \label{eqn: Asymp-var}
\asv = \frac{\left[\frac{\sigv}{\powcnst}+1-\Chardeta\right]}{2 \w^2 \Charseta}
\end{equation}
\end{thm}

\begin{IEEEproof}
Please see Appendix 1.
\end{IEEEproof}

Note that in the i.i.d. case (\ref{emp}) is the empirical characteristic function (ECF) \cite{ushakov1999} of
$\ieta$ corrupted by additive noise. While the ECF has been studied extensively in the
statistical literature for constructing centralized estimators \cite{ushakov1999},
it has not been addressed in the context of communication of samples as in distributed estimation,
and therefore issues of power constraint and channel noise have not arisen in the literature on parameter
estimation with ECFs.

\section{Analysis and Optimization of the $AsV$} \label{sec: analysis_n_opt_AsV}

The proposed estimator is consistent under general conditions and does not depend on the noise parameters.
However, if the noise distribution and parameters are available, it is possible to minimize the $AsV$ with respect to $\w$ over the interval $(0, 2\pi/\thetarng]$:
\begin{equation}\label{eqn: opt_asymp_var}
{AsV^{\ast}}:=\inf_{\w \in \left(0, 2\pi/\thetarng\right]} \frac{\left[\frac{\sigv}{\powcnst }+1-\Chardeta \right]}{2\w^2\Charseta}.
\end{equation}
We will consider this problem with both per-sensor, and total power constraints.

\subsection{Per-sensor Power Constraint}\label{sec: per_sensor_pow_cnst}


Our derivation for the estimator ${\hat \theta}$ in (\ref{eqn: prop_estimator}), its strong consistency
in Theorem \ref{thm1}, and the asymptotic variance in (\ref{eqn: Asymp-var}) had assumed that  $\powcnst$ is fixed
as a function of $L$.
In the fixed per-sensor power constraint case the total power $\powcnst= \rho L$ increases
linearly with $L$ in which case the estimator is given in (\ref{eqn: prop_estimator}) with $z_L:= y_L/L$ which we redefine with an extra
factor of $1/\sqrt{L}$ in (\ref{emp}).
In this case, the statement of Theorem \ref{thm1} still holds exactly, with minor modifications in the proof, and
$\sigv/\powcnst  \rightarrow 0$ as $L \rightarrow \infty$. Hence, having a per-sensor
power constraint is asymptotically equivalent to having no channel noise. In either case (\ref{eqn: opt_asymp_var}) becomes,
\begin{equation} \label{eqn: asymp_var3}
AsV^{\ast}_{\rm pspc} = \inf_{\w \in \left(0, 2\pi/\thetarng\right]} \frac{\left[1-\Chardeta \right]}{2 \w^2 \Charseta},
\end{equation}
which is a special case of (\ref{eqn: opt_asymp_var}). The reason we consider this case separately is
because, as we will see, the objective in (\ref{eqn: asymp_var3}) is bounded near the origin
which makes the solution of (\ref{eqn: asymp_var3}) considerably different than that of (\ref{eqn: opt_asymp_var}).
We now consider solving (\ref{eqn: asymp_var3}), and investigate the behavior of $\asv$ near
the origin to see under what conditions small $\w$ will yield optimum performance.
Using l'H\^{o}spital's rule, it is seen that
$\lim_{\w \rightarrow 0} \asv = \sigeta$ the variance of $\eta$, when $\eta$
has finite variance. In fact, when also the fourth moment $\mu_4$ of $\eta$ exists, we have a stronger result:

\begin{thm}\label{thm3}
If the first four moments of $\eta$ exists, then $\asv$ in equation (\ref{eqn: asymp_var3}) satisfies
\begin{equation}\label{eqn: asymp_var4}
\asv = \sigeta-\frac{1}{3} \kurt \sigma_{\eta}^4\w^2+o(\w^2)
\end{equation}as $\w \rightarrow 0$, where $\kurt : = \mu_4/\sigma_{\eta}^4-3$ is the excess kurtosis of $\eta$.
\end{thm}

\begin{IEEEproof}
We have already established that the first term in (\ref{eqn: asymp_var4}) is $\sigeta$.
Using the Maclaurin series expansion of $\Chareta$ in terms of the second and fourth moments of $\eta$, the numerator and denominator of (\ref{eqn: asymp_var3}) can be expressed as $N(\w) := 2 \sigeta\w^4+(2/3)\mu_4\w^4+o(\w^4)$ and $D(\w):= 2 \w^2 \left(1-(1/2)\sigeta\w^2+(\mu_4/24)\w^4+o(\w^4)\right)$, respectively. By taking the second derivative and evaluating we have
\begin{equation} \label{exp}
\frac{\partial^2}{\partial \w^2}\frac{N(\w)}{D(\w)}\bigg|_{\w=0} = -\frac{2}{3}\mu_4+2\sigma_{\eta}^4.
\end{equation}
Dividing by $2!$ we obtain the coefficient of $\w^2$ in the Maclaurin series, as given in (\ref{eqn: asymp_var4}).
\end{IEEEproof}

Theorem \ref{thm3} has some interesting implications. By making $\w$ sufficiently small, we
can obtain an $AsV$ that is arbitrarily close to $\sigeta$. Also, if the excess kurtosis
of the sensing noise is positive, it is possible to improve the $AsV$ to a value smaller than $\sigeta$
by increasing $\w$ in the neighborhood of $0$, which shows that if $\kurt >0$, (\ref{eqn: asymp_var3}) satisfies
$AsV^{\ast}_{\rm pspc} < \sigeta$.
This is the case for impulsive distributions like the Laplace distribution where $\kurt = 3$.
When $\eta$ is Gaussian, the excess kurtosis $\kurt=0$ and therefore it is not clear from (\ref{eqn: asymp_var4}) if
$AsV^{\ast}_{\rm pspc} < \sigeta$ is possible, since (\ref{eqn: asymp_var4}) only applies near $\w=0$.
The following theorem sheds more light on this issue.
\begin{thm}\label{thm3.5}
If $\eta$ is Gaussian then the best asymptotic performance for ${\hat \theta}$ in (\ref{eqn: prop_estimator}) for the per-sensor power
constraint satisfies $AsV^{\ast}_{\rm pspc} = \sigeta$.
\end{thm}
\begin{IEEEproof}
Equation (\ref{eqn: asymp_var4}) shows that $\lim_{\w \rightarrow 0} \asv = \sigeta$ which implies that
$AsV^{\ast}_{\rm pspc} \leq \sigeta$.
To see that $AsV^{\ast}_{\rm pspc} \geq \sigeta$
consider a benchmark genie-aided sample mean estimator $\esttheta_{GA} = L^{-1} \sum_{i=1}^{L} \isrx$ that has access to
the sensor measurements  $\{\isrx\}_{i=1}^L$, rather than the the normalized channel output $z_L$ in (\ref{emp}).
The sample mean which has an asymptotic variance of $\sigeta$
achieves the Cramer Rao bound (CRB) for an estimator of $\theta$
from $\{\isrx\}_{i=1}^L$ since it is an
efficient estimator of the mean when $\eta$ is Gaussian.
Since $\theta \rightarrow \{\isrx\}_{i=1}^L \rightarrow z_L$ forms a Markov chain,
from the data processing inequality for the CRB \cite{zamir}, the CRB
for estimators of $\theta$ based on $z_L$ 
is at least that obtained for the genie-aided setup of estimating $\theta$ from $\{\isrx\}_{i=1}^L$,
which is $\sigeta$.
Therefore, the best achievable performance in the per-sensor power case cannot
be better than that of
${\hat \theta}_{\rm GA}$, which implies $AsV^{\ast}_{\rm pspc}\geq \sigeta$.
\end{IEEEproof}

Note that in the proof of Theorem \ref{thm3.5} we used the Gaussianity only to assert that the
sample mean achieves the CRB. Theorem \ref{thm3.5} also holds for any other distribution with this
property.

In Figure \ref{fig: asv_w_psc} we show the analytical expressions for $AsV(\w)$ for various distributions.
For the Cauchy distribution the performance is unbounded at the origin since the variance does
not exist. For all other distributions, we selected $\sigeta =1$, which is the value of $AsV(\w)$ near
the origin. Note that the Laplace distribution which
has a positive excess kurtosis corresponds to an $AsV$ which is decreasing  near the origin, as predicted by (\ref{eqn: asymp_var4}), whereas
the Gaussian and uniform distributions are increasing near the origin from their infimum value of $\sigeta =1$.

To conclude, for this per-sensor power constraint case, small $\w$ yields good asymptotic performance
which does not depend on $\theta_R$. The performance
can be improved by appropriately increasing $\w$ in the neighborhood
of $\w=0$ when $\eta$ is from an impulsive distribution with positive excess kurtosis.

\subsection{Total Power Constraint} \label{sec: total_pow_cnst}

In this case $\powcnst $ is not a linear function of $L$ as it was in the previous section, but a
constant so that
the $AsV$ is given by (\ref{eqn: Asymp-var}).
Note that small $\omega$ should be avoided in the solution of (\ref{eqn: opt_asymp_var})
since $\lim_{\w \rightarrow 0} AsV(\w) = \infty$ is no longer finite, as seen also in Figure \ref{fig: asv_w_tpc} for various
fading distributions.
For the same reason, one may use $\min$ instead of $\sup$ in (\ref{eqn: opt_asymp_var}) for the total
power constraint case, since the minimum is always achieved by a strictly positive $\w$, when $\sigma_v^2>0$.
\subsubsection{Upper bound on AsV}
In what follows, we use the lower bound $\Chareta \geq 1-\sigeta\w^2/2$
in order to upper bound $AsV^{\ast}$ in (\ref{eqn: opt_asymp_var}).
We have the following theorem which applies when $\thetarng$ is large enough so that $(2\pi \sigma_{\eta})/\thetarng<\sqrt{2}$.

\begin{thm}\label{thm4}
The best achievable performance $AsV^{\ast}$ in (\ref{eqn:
opt_asymp_var}) for any sensing noise distribution with finite
variance  $\sigeta$ satisfies
\begin{equation}\label{eqn: ineq1}
AsV^{\ast} \leq \frac{\left[\frac{4\sigv}{\powcnst}+c\right]}{c\left[1-\frac{1}{16}c\right]^2}
\end{equation}whenever $c/8<(2\pi/\thetarng)^2 \sigeta<2$, where $c := -3\sigv/\powcnst+(\sigma_v/\sqrt{\powcnst})\sqrt{32+9 \sigv/\powcnst}$.
On the other hand, when $(2\pi/\thetarng)^2 \sigeta <c/8$ then,
\begin{equation}\label{eqn: ineq2}
AsV^{\ast} \leq \frac{\left[\frac{\sigv}{\powcnst}+2\left(\frac{2\pi}{\thetarng}\sigma_{\eta}\right)^2 \right]}{2\left(\frac{2\pi}{\thetarng}\right)^2 \left(1-\left(\frac{2\pi}{\thetarng}\sigma_{\eta} \right)^2 \frac{1}{2} \right)^2}
\end{equation}

\end{thm}

\begin{IEEEproof} Please see Appendix 2.
\end{IEEEproof}

Note that if the range of the unknown parameter determined by $\thetarng$ is large,
the upper bound in (\ref{eqn: ineq2}) will be tighter since the bound $\Chareta \geq 1-\sigeta\w^2/2$
is tighter when $\w$ is small. Moreover, when $\thetarng$ is large, (\ref{eqn: ineq2}) simplifies
to $(\sigv/\powcnst )(\theta_R^2/8\pi^2) + \sigeta$
This shows that if the range $\thetarng$
increases, the optimal achievable performance $AsV^{\ast}$ increases as well.
In addition to large $\theta_R$, when $\sigma_{v}^2/P_T \rightarrow 0$ as in the per-sensor
power constraint case, the bound further simplifies to $AsV^{\ast} \leq \sigeta$.
The bound in Theorem \ref{thm4} holds regardless of the distribution
on $\eta$ as long as it has finite variance.

Instead of working with bounds, if exact solutions to (\ref{eqn: opt_asymp_var}) are desired, then
it is necessary to specify the sensing noise distribution. In what follows, the problem is
specialized by considering some common distributions. The resulting asymptotic variances for the different
distributions are illustrated in Figure \ref{fig: asv_w_tpc}.


\subsubsection{Gaussian Sensing Noise}\label{sec: gauss_sens_noise}

In this case, we have $\Chareta = \exp\left(-\sigeta\w^2/2\right)$ so that
\begin{equation}\label{eqn: asymp_var_gauss}
AsV_{\rm G}(\w) = \frac{e^{\sigeta\w^2}}{2
\w^2}\left[\frac{\sigv}{\powcnst }+1-e^{-2 \sigeta \w^2} \right].
\end{equation}

We would like to minimize (\ref{eqn: asymp_var_gauss}) over $\w \in (0,2 \pi/\theta_R]$ as in
(\ref{eqn: opt_asymp_var}).
As an intermediary step,
we first characterize the unconstrained minimum over $\w \in [0, \infty)$.
To simplify (\ref{eqn: asymp_var_gauss}) we substitute $\bt \leftarrow \sigeta\w^2$.
Note that the value of $\w$ that minimizes (\ref{eqn: asymp_var_gauss}) over $\w >0$
is related to the $\beta>0$ that minimizes $AsV_{\rm G}(\sqrt{\beta}/\sigma_{\eta})$
through $\w=\sqrt{\beta}/\sigma_{\eta}$.
Differentiating with respect to $\bt$, we have,
\begin{equation} \label{eqn: diff_asymp_var_beta}
\frac{\partial AsV_{\rm G}(\sqrt{\bt}/\sigma_{\eta})}{\partial \bt} =
\frac{e^{-\bt}\left[\left(\frac{\sigv}{\powcnst
}+1\right)\left(\bt-1 \right)e^{2\bt}+\bt+1 \right]\sigeta}{2\bt^2}
\;.
\end{equation}Any stationary point of $AsV_{\rm G}(\sqrt{\bt}/\sigma_{\eta})$, with respect to $\bt$ satisfies,
\begin{equation}\label{eqn: stat_pnt_asymp_var}
\left(\frac{\sigv}{\powcnst}+1 \right)\left(\bt-1 \right)e^{2\bt}+\left(\bt+1 \right) = 0 \;.
\end{equation}
Let any solution to (\ref{eqn: stat_pnt_asymp_var}) be denoted as
$\betasol$. It is straightforward to show that
$\partial^2 AsV_{\rm G}(\sqrt{\beta}/\sigma_\eta)/\partial \beta^2|_{\beta=\betasol}$ is positive.
This proves that $\betasol$ is the unique unconstrained minimum of
$AsV_{\rm G}(\sqrt{\bt}/\sigma_{\eta})$ over $\beta>0$ which
in turn implies that $\w_{\rm G}^* = \sqrt{\betasol}/\sigma_\eta$ is the corresponding
unique minimizer of
$AsV_{\rm G}(\w)$ for $\w>0$.
Since $AsV_{\rm G}(\w)$ has a unique minimum, it is monotonically
decreasing over $\w \in (0, \sqrt{\betasol} / \sigma_{\eta}]$. The solution
to (\ref{eqn: opt_asymp_var}) in the Gaussian case therefore is
\begin{equation}\label{eqn: guass_omega}
\w^{\ast}_G = \min\left(\frac{2\pi}{\thetarng}, \frac{\sqrt{\betasol}}{\sigma_{\eta}}\right),
\end{equation}where $\betasol$ is the unique solution to (\ref{eqn: stat_pnt_asymp_var}).

While there has been some efforts in the physics community \cite{scott2006} to define functions that
solve the intersection point of rational functions and exponentials as in (\ref{eqn: stat_pnt_asymp_var}),
there is no widely accepted formula. But (\ref{eqn: stat_pnt_asymp_var}) can be easily solved numerically
to optimize $\w$ when $\eta$ is Gaussian.

\subsubsection{Cauchy Sensing Noise}\label{sec: cauchy_sens_noise}

For the Cauchy distribution $\Chareta = e^{-\gm \w}$ for $\w>0$. It is well known that no moments of this distribution exists. Substituting $\Chareta$ in (\ref{eqn: Asymp-var}), we have

\begin{equation}\label{eqn: Cauchy_asymp_var}
AsV_{\rm C}(\w) = \frac{e^{2 \gm \w}}{2\w^2}
\left[\frac{\sigv}{\powcnst }+1-e^{-2\gm \w} \right] .
\end{equation}As in the Gaussian case, we first find the stationary points of (\ref{eqn: Cauchy_asymp_var}) on $\w \in [0, \infty)$ by taking the derivative of (\ref{eqn: Cauchy_asymp_var}) and equating to zero to obtain,
\begin{equation}\label{eqn: caucy_beta}
\beta_C^{\ast} = \frac{1}{2\gm} \left[2+W\left(\frac{-2\powcnst }{\sigv+\powcnst }e^{-2}\right)\right]
\end{equation}where $W(\cdot)$ is the Lambert function defined to be the inverse function of $xe^{x}$. It can be verified that $AsV''(\beta_C^{\ast})>0$ and therefore $\beta_C^{\ast}$ is the unique unconstrained minimum of $AsV_{\rm C}(\w)$. Hence, $AsV_{\rm C}(\w)$ has a unique minimum
over $\w >0$, and the solution to (\ref{eqn: opt_asymp_var}) in this case is
\begin{equation}\label{eqn: cauchy_omeg}
\w^{\ast}_C = \min \left(\frac{2\pi}{\thetarng}, \beta_C^{\ast} \right).
\end{equation}

\subsubsection{Laplace Sensing Noise}\label{sec: lapl_sens_noise}

In this case, we have $\Chareta = (1+b^2\w^2)^{-1}$ where $b^2 :=
\sigeta/2$. Substituting $\beta \leftarrow b^2\w^2$ for convenience,
(\ref{eqn: Asymp-var}) for Laplace noise becomes,

\begin{equation}\label{eqn: asymp_laplace}
AsV_{\rm L}\left(\sqrt{\frac{\beta}{b^2}}\right) = b^2
\frac{(1+\beta)^2}{2\beta} \left[\frac{\sigv}{\powcnst
}+\frac{4\beta}{1+4\beta} \right].
\end{equation}

To characterize the stationary points of (\ref{eqn: asymp_laplace}), we take the derivative with respect to $\beta$ and equate to zero. The optimum value is the root of a $4^{th}$ order polynomial. Using the only solution with a positive root we have,

\begin{eqnarray}
\beta^{\ast}_L = \frac{1}{12} \left( \frac{{c}}{\frac{\sigv}{\powcnst }+1}+\frac{25 \frac{\sigv}{\powcnst }+4}{{c}}+2 \right) \label{eqn: laplace_beta}
\end{eqnarray}
where
\begin{eqnarray}
c=\left[125\left(\frac{\sigv}{\powcnst }\right)^3{+}258\left(\frac{\sigv}{\powcnst }\right)^2{+}141\left(\frac{\sigv}{\powcnst }\right) {+} 3\sqrt{3}\sqrt{\left(\frac{\sigv}{\powcnst }\right)\left(\frac{\sigv}{\powcnst }{+}1\right)^3\left(375\frac{\sigv}{\powcnst }{+}32\right)}{+}8 \right]^{1/3}. \nonumber
\end{eqnarray}

It is also possible to verify that the second derivative is positive
at the optimal point. To express the roots of the $4^{th}$ order
polynomial in closed-form and verifying that the second derivative
is always positive, 
we have used ${\text{Mathematica}}$. Using (\ref{eqn:
laplace_beta}), $\w^2 b^2 = \beta$, and the fact that $\w \in
(0,2\pi/\thetarng]$ we have the solution to (\ref{eqn: opt_asymp_var}) as

\begin{equation}\label{eqn: opt_omega_lapl}
\w^{\ast}_L = \min\left(\frac{2\pi}{\thetarng}, \frac{1}{b}\sqrt{\beta^{\ast}_L} \right).
\end{equation}

\subsubsection{Uniform Sensing Noise}\label{sec: uni_sens_noise}

We now assume that $\eta$ is uniformly distributed on $[-a, \hspace{0.05 in} a]$, so that $a^2=3\sigeta$.
In this case $\Chareta = \sin(\w a)/(\w a)$ and we need to optimize,
\begin{equation} \label{unif}
AsV_{\rm U}(\w) = \frac{a^2}{2\sin^2(\w a)}
\left[\frac{\sigv}{\powcnst }+\left(1-\frac{\sin(2\w a)}{2\w a}
\right)\right]
\end{equation}over $\w \in(0, 2\pi/\thetarng]$.
Note that $AsV_{\rm U}(\w)$ is undefined at $\w=\pi/a$. We begin by showing that the range of
$\w$ can be further reduced to
$\w \in (0, \min(2\pi/\thetarng, \pi/a))$ in solving (\ref{eqn: opt_asymp_var}). This is because
both $\sin^2(\w a)$ and $\sin(2\w a)$ are periodic with period $\pi/a$, and therefore due to the
$2\w a$ term in the denominator,
$AsV_{\rm U}(\w) \leq AsV_{\rm U}(\w+k\pi/a)$ for any positive integer $k$, and $\w >0$.

In order to minimize (\ref{unif}) over $\w \in (0, \min(2\pi/\thetarng, \pi/a))$, we first
disregard the constraint on $\w$ imposed by $\thetarng$, and focus on $\w \in (0,\pi/a)$. Substituting $\beta \leftarrow \w a$, differentiating
$AsV_{\rm U}(\beta/a)$ with respect to $\beta$ and equating to zero we obtain
\begin{eqnarray} \label{unifsoln}
\left[8 \left(\frac{\sigma_v^2}{P_T}+1\right) \beta^2-1\right]
\cos(\beta) + \cos(3\beta) - 4 \beta \sin(\beta) = 0 \;.
\end{eqnarray}
By taking the second derivative, it can be verified that of $AsV_{\rm U}(\beta/a)$ is convex, and therefore (\ref{unifsoln}) has
a unique solution $\beta_{\rm U}^*$ over $\beta \in (0,\pi)$ corresponding to the unique minimum of
$AsV_{\rm U}(\beta/a)$ over the same interval. It is immediate that $\w = \beta_{\rm U}^*/a$ is the
unique minimum of (\ref{unif}) over $\w \in (0,\pi/a)$, and therefore (\ref{unif}) is a monotonically
decreasing function over $(0,\beta_{\rm U}^*/a)$. 
Incorporating the effect of $\theta_R$, we have that if $2\pi/\thetarng \leq \beta_{\rm U}^*/a$
then the minimum of (\ref{unif}) over $\w \in (0, \min(2\pi/\thetarng, \pi/a))$ is attained at
$\w=2 \pi/\theta_R$, and if $2\pi/\thetarng \geq \beta_{\rm U}^*/a$, then it is attained at $\w=\beta_{\rm U}^*/a$.
In short,
\begin{eqnarray} \label{ounif}
\w^*_{\rm U} = \min\left(\frac{2 \pi}{\theta_R},\frac{\beta^*_{\rm U}}{a}\right)
\end{eqnarray}
Note that a closed-form solution to (\ref{unifsoln}) is not possible, however a numerical solution can
be easily found. Recall also from Section \ref{sec: analysis_n_opt_AsV} that for uniform noise which has $\kurt = -6/5$, a small $\w>0$ should be chosen when $\sigv/\powcnst =0$. If instead $\sigv/\powcnst >0$, then $\w \approx \pi/2a$ (or $2\pi/\thetarng$, whichever one is smaller) is a good choice. We will elaborate on this more in Section \ref{sec: low_snr_regime}, where we consider the low channel SNR regime.

\subsubsection{Compound Gaussian Sensing Noise}\label{sec: cmpd_gauss_sens_noise}

Compound Gaussian is a class of RVs which when conditioned on the variance is a Gaussian RV. So when $\eta$ is compound Gaussian, it can be written as $\eta=\sqrt{X} G$ where $G$ is
a Gaussian RV with zero mean and variance one, and $X$ is a positive RV.
It is easy to show that the CF of $\eta$ can be expressed in terms of the
moment generating function (MGF) of $X$:
\begin{equation}\label{eqn: cmpd_gauss_MGF}
\Chareta = {\rm E} \left[e^{-\frac{1}{2}X\w^2}\right] = M_X\left(-\frac{1}{2} \w^2\right)
\end{equation}
where $M_X(t):= {\rm{E}}[e^{tX}]$ is the MGF of $X$ when the expectation exists. Note that ${\rm{E}}[X] = \sigeta$ in general and if the CDF of $X$ is a unit step at $\sigeta$ then $\eta$ is Gaussian with variance $\sigeta$.
For compound Gaussian sensing noise, (\ref{eqn: cmpd_gauss_MGF}) can be substituted in (\ref{eqn: Asymp-var}) to obtain
\begin{equation}\label{eqn: cg_asymp_var}
AsV_{\rm CG}(\w) = \frac{\left[\frac{\sigv}{\powcnst}+1-M_X(-2\w^2)\right]}{2\w^2M_X^2(-\frac{1}{2}\w^2)}
\end{equation}whenever the MGF exists.

When the per-sensor power is fixed so that $\sigv/P_{T} \rightarrow 0$ as $L \rightarrow \infty$, (\ref{eqn: cg_asymp_var}) can be expanded near $\w = 0$ to obtain,
\begin{equation}\label{eqn: cg_asymp_var_near_zero}
AsV_{\rm CG}(\w) = {\rm{E}}[X]-\sigx  \w^2+o(\w^2)
\end{equation}which is the same as (\ref{eqn: asymp_var4}), expressed in terms of the mean and variance of $X$. When $\sigx  = 0$, $X$ is a constant and $\eta$ is Gaussian. If instead $\sigx  >0$, then $AsV$ can be improved by increasing $\w$ in the neighborhood of 0, implying that
$AsV^\ast_{\rm pspc} < \sigeta$.

\begin{figure}[tb]
\begin{minipage}{1\textwidth}
\centering
\begin{center}
\includegraphics[height=9cm,width=12cm]{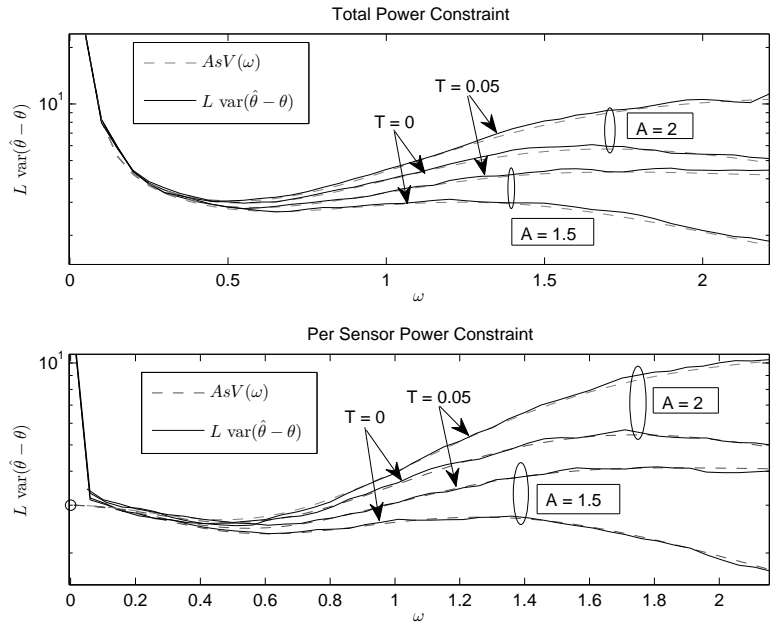}
\caption{$\sigma_{\eta}^2 = 3$, $\sigma_v^2 = 1$, $L = 500$, $P_T = 10$}\label{fig: cmpd_gauss_w}
\end{center}
\end{minipage}
\end{figure}

As a concrete example, consider Middleton Class-A noise \cite{poor89} where the variance RV is discrete and given by $X = \sigeta\left[Y/(A(T+1))+T/(T+1) \right]$, $A$ and $T$ are deterministic
parameters controlling the impulsiveness of the noise $\eta$, and $Y$ is a Poisson RV with parameter $A$. In this case,

\begin{equation}
M_X(t) = \exp\left(t\frac{\sigeta T}{T+1}\right)\exp\left(A \left(\exp\left(\frac{t\sigeta}{A(T+1)}\right)-1 \right) \right) \;.
\end{equation}
Substituting in (\ref{eqn: cg_asymp_var}) we obtain the $AsV$. The resulting expression shows that when $T=0$ (highly impulsive noise) $AsV_{\rm CG}(\w) \rightarrow 0$ as $\w \rightarrow \infty$ in which case $\w$ should be chosen as large as possible (i.e., $\w = 2\pi/\thetarng$). Another interesting aspect of this expression is that it illustrates that $AsV(\w)$ need not have a unique local minimum (i.e., it need not be convex or quasi-convex) for
every sensing noise distribution. In fact, as will be seen in Figure \ref{fig: cmpd_gauss_w} of the Simulations section,
$AsV(\w)$ can have multiple local minima, unlike the Gaussian, Cauchy and Laplace cases considered thus far.

\subsection{Low Channel SNR Regime}\label{sec: low_snr_regime}

When $\sigv/\powcnst $ is sufficiently large, the $\Chardeta$ term in (\ref{eqn: Asymp-var}) is negligible, thereby transforming the problem in (\ref{eqn: opt_asymp_var}) into maximizing $[\w \Chareta]^2$ over $(0, 2 \pi/\thetarng]$. We now briefly summarize how the solutions in the previous subsection simplify in this regime. Since we already have closed form expressions for the solution of (\ref{eqn: opt_asymp_var}) for the Cauchy and Laplace cases, we only focus on the Gaussian and uniform cases.

For the Gaussian case maximizing $\w^2 e^{-\sigeta \w^2}$ over $\w \in (0, 2\pi/\thetarng]$ yields $\w^{\ast} = \min \left(2\pi/\thetarng, 1/\sigma_{\eta} \right)$. If $\thetarng$ is sufficiently small so that $\w^{\ast} = 1/\sigma_{\eta}$, then we have
\begin{equation}\label{ubbap}
AsV_{\rm G}(1/\sigma_{\eta}) = \frac{\sigeta
e}{2}\left[\frac{\sigv}{\powcnst }+(1-e^{-2})  \right]
\end{equation} which is an upper bound on the best achievable performance $AsV^{\ast}$, even when
the channel SNR is not low, but becomes tighter at low channel SNR.

For the uniform case we maximize $\sin^2(\w a)$ which yields $\w^{\ast} = \min \left(2\pi/\thetarng, \pi/2a  \right)$. If $\thetarng$ is
small enough, $\w^{\ast} = \pi/(2a)$ and $AsV_{\rm U}(\pi/2a) = (a^2/2)(\sigv/\powcnst)$.

\section{Comparison with Amplify and Forward Scheme}\label{sec: comp_a_n_f}

In the AF scheme, the transmitted signal at the $i^{th}$ sensor is $\alpha_L \isrx$ where $\alpha_L$ depends on
the number of sensors $L$ to maintain the total power constraint, but is independent of $\isrx$ \cite{cui2007}, \cite{gastparb2003}.
We focus on the i.i.d. case for simplicity, and choose $\alpha_L$ identical across sensors due to symmetry. In what follows,
we will show that the asymptotic performance of AF is competitive with that of the proposed scheme when
the sensing noise has finite variance, and inferior to the proposed scheme when the sensing noise is impulsive.

The received signal for AF is,
\begin{equation}\label{eqn: fusion_rx_AF}
y_L = \alpha_L \sum_{i=1}^L (\ttheta+\ieta)+v\;.
\end{equation}
We have already alluded to the fact that the per-sensor power $\alpha_L^2 (\ttheta+\ieta)^2$ is an unbounded RV,
when the pdf of the sensing noise has infinite support. This is undesirable especially for low-power sensor networks with limited peak-power capabilities. Therefore, before we compare the asymptotic variances of the proposed estimator and AF, we reiterate that with respect to the management of the instantaneous transmit power of sensors, the proposed estimator is preferable to AF.

Since the total instantaneous power is random for AF, the total power is defined as an average $\powcnst  = \alpha_L^2 \sum_{i=1}^{L} {\rm E}[(\ttheta+\ieta)^2]$, with respect to the sensing noise distribution.
We will consider a total power constraint case where $\powcnst$ is not a function of $L$ so that
$\alpha_L = \sqrt{\frac{\powcnst}{L(\theta^2+\sigeta)}}=O(L^{-1/2})$.


The estimator in AF is given by $\esttheta_{AF} = {\rm
Re}\{\frx\}/(L \alpha_L)$ so that
\begin{equation} \label{asvaf}
\sqrt{L} (\esttheta_{AF} - \theta) = \frac{1}{\sqrt{L}} \sum_{i=1}^L
\ieta + \sqrt{\frac{\theta^2+\sigeta}{\powcnst}} \; {\rm Re}\{v\}
\end{equation}
with an $AsV$ of,
\begin{equation}\label{eqn: asymp_var_AF}
AsV_{\rm AF} = \sigeta +\frac{\sigv}{2\powcnst}(\ttheta^2+\sigeta)
\end{equation}
when $\eta$ has finite variance.

Consider now the special case of no channel noise ($\sigv = 0$)
which implies $AsV_{\rm AF} = \sigeta$. In Section \ref{sec:
per_sensor_pow_cnst} we have seen that $AsV^{\ast}_{\rm pspc} <
\sigeta$ is possible when the sensing noise is impulsive enough to
have a positive excess kurtosis, the proposed approach outperforms
AF when there is no channel noise. We now examine the more general
case of $\sigv>0$.

Observe that (\ref{eqn: asymp_var_AF}) depends explicitly on
$\ttheta$, whereas (\ref{eqn: opt_asymp_var}) depends on the
estimation range $\theta_R$. Since it is difficult to compare these
expressions in general, we will examine the case of large and small
$\ttheta$. When $\ttheta$ is large, $AsV_{\rm AF} \approx \sigeta +
(\sigv/\powcnst ) (\ttheta^2/2)$, and by the discussion after
(\ref{eqn: ineq2}), $AsV^{\ast} \approx \sigeta + (\sigv/\powcnst
)(\theta_R^2/8\pi^2)$.
Note that when the parameter $\theta$ is close to its upper limit, the proposed estimator will
outperform AF. However, when $\theta$ is very small despite a large
range $\theta_R$, the AF will outperform the proposed approach.

Let us now examine the case of small $\theta$ and $\theta_R$, where we focus on the Gaussian
case. For this purpose, we bound the difference in performance between the proposed estimator
and AF:
\begin{eqnarray}
AsV^{\ast}-AsV_{\rm AF}
&\leq&
AsV_{\rm G}(1/\sigma_\eta) - AsV_{\rm AF} \\
&=& \label{bd2} \sigeta \left[\frac{e}{2}(1-e^{-2})-1
+\frac{\sigv}{2\powcnst} \left(e-1-\frac{\theta^2}{\sigeta}\right)
\right]\;,
\end{eqnarray}
where the inequality is because (\ref{ubbap}) is an upper bound on
$AsV^{\ast}$. Examining the bound in (\ref{bd2}) we note that its
sign depends on the the channel SNR $\sigv/\powcnst$ and the sensing
SNR $\theta^2/\sigeta$. In conclusion, the proposed approach is
competitive with AF and may outperform it, depending on the specific
parameter values when the sensing noise has finite variance. In what
follows, the heavy-tailed sensing noise case is discussed.

With the AF approach the normalized multiple access channel output
is proportional to the the sample mean, which is not a good estimator of $\theta$ when the
sensing noise is heavy-tailed. To illustrate with a specific
example, consider the case when $\eta$ is Cauchy. Dividing both
sides of (\ref{asvaf}) with $\sqrt{L}$ it is clear that
$(\esttheta_{AF} - \theta) \rightarrow 0$ is not possible since the
sample mean  $L^{-1}\sum_{i=0}^L \ieta$ is Cauchy distributed and
has the same distribution as $\eta_1$ regardless of the value of $L$. Since the sample mean is not
a consistent estimator for Cauchy noise, the AF approach over
multiple access channels fails for such a heavy-tailed distribution.
On the other hand, the proposed estimator is strongly consistent in
the presence of any noise distribution, including Cauchy. This brief
example illustrates that the inherent robustness of our approach in
the presence of heavy-tailed sensing noise distributions. The
sample mean, ``computed'' by the multiple access channel in the
AF approach, is highly suboptimal, and sometimes not consistent like in the Cauchy case,
whereas in the proposed approach
the channel computes (a noisy and normalized version of) the empirical characteristic
function of the sensed samples, from which a consistent estimator can be constructed
for any sensing noise distribution.

To be fair to AF, even though it suffers from having potentially large
peak powers, we also want to point out the situations under which
it is preferable to the proposed approach. The first point is that
AF does not require the parameter $\theta$ to be bounded, and it does
not require fine-tuning of a transmission parameter like $\w$. Moreover,
AF is also a ``universal'' estimator, albeit over a smaller class of
distributions (those that have finite variance) for the sensing noise.

In conclusion, the proposed estimator with its fixed instantaneous
power per sensor is inherently preferable to AF when the sensors
have a small dynamic range. Moreover, for AF, the total transmit power depends
on $\theta$ and the statistics of the sensing noise.
On the other hand, the AF approach has the benefit of not
assuming $\theta$ to be in a finite set, and sometimes has a better finite sample
performance as seen in the simulations. For impulsive noise
distributions with finite variance and positive excess kurtosis like
Laplace, or heavy-tailed distributions with infinite variance like
Cauchy, the proposed approach is superior to AF. For other regimes,
the two schemes are competitive and their asymptotic performance
comparison depends on the specific values of parameters $\theta$,
$\theta_R$ $\sigv$, $\sigeta$, and $\powcnst$.

\section{Fading Channels}

Suppose that the multiple access channel connecting the sensors to the FC has fading so that (\ref{eqn: fusion_cntr_rx}) becomes
\begin{equation}\label{eqn: fusion_center_rx_fading}
\frx = \sqrt{\rho} \sum_{i=1}^{L} |h_i| e^{j\w \isrx}+\feta
\end{equation}where $|h_i|$ is the amplitude of the channel coefficient $h_i$ between the $i^{th}$ sensor and the FC satisfying E$[|h_i|^2]=1$.
Even though the channel $h_i$ is complex valued, the effective channel $|h_i|$ is real and positive when the $i^{th}$ sensor corrects for the channel
phase before transmission, using local channel phase information. Such a phase correction does not change the constant power nature of the transmission.

The following Theorem characterizes the performance of the proposed estimator over fading channels:

\begin{thm}\label{thm5}
For the channel in (\ref{eqn: fusion_center_rx_fading}) the estimator $\esttheta$ in (\ref{eqn: prop_estimator}) is asymptotically normal with variance
\begin{equation}\label{eqn: asymp_var_fading}
AsV(\w) = ({\text{E}}[|h_i|])^{-2} \frac{\left[\frac{\sigv}{\powcnst}+1-\Chardeta\right]}{2\w^2 \Charseta}
\end{equation}
\end{thm}

\begin{IEEEproof}
The proof is similar to that of Theorem \ref{thm1} with the following changes:
\begin{eqnarray}
v_c &=& \frac{1}{2}+\frac{1}{2}\Chareta-({\text{E}}[|h_i|])^2 \Charseta \nonumber \\
v_s &=& \frac{1}{2}-\frac{1}{2}\Chardeta \nonumber ,
\end{eqnarray}and both $G_1$ and $G_2$ are scaled by a factor of $({\text{E}}[|h_i|])^{-1}$. Substituting these in (\ref{eqn: asymp_var2}) we obtain (\ref{eqn: asymp_var_fading}).
\end{IEEEproof}

Since ${\text{E}}[|h_i|^2]=1$, using Jensen's inequality, the $({\text{E}}[|h_i|])^{-2}$ factor due to fading is always less than
one, unless $|h_i|$ is deterministic. In fact, when $|h_i|$ is Ricean the loss due to fading is given by
$(\sqrt{K+1}~\Gamma(3/2) e^{-K}~ _{1}F_{1}(3/2;1; K))^{-2}$
where $\text{ }_{1}F_{1}(\cdot;\cdot;\cdot)$ is the confluent
hypergeometric function \cite[pp. 504]{Abra65} and $K$ is the Ricean parameter. This expression reduces to $4/\pi$ when $K=0$, implying Rayleigh
fading channels. In the AF setting, the difference between fading and no fading also exhibits the same loss,
which was analyzed in detail in \cite{maheshicassp,maheshasilomar} for different fading distributions, where the Nakagami case was also considered.
Note that if the optimization of the asymptotic variance is desired in the fading case, the fading loss does not affect the optimum value of $\w$ so
equations (\ref{eqn: guass_omega}), (\ref{eqn: cauchy_omeg}), (\ref{eqn: opt_omega_lapl}), and (\ref{ounif}) remain valid for the different sensing noise distributions.

\section{Simulations}
In what follows, we corroborate our analytical results through Monte Carlo simulations, and also
examine finite-sample effects that are not predictable from our asymptotic results.

\begin{figure}[tb]
\begin{minipage}{1\textwidth}
\centering
\begin{center}
\includegraphics[height=9cm,width=12cm]{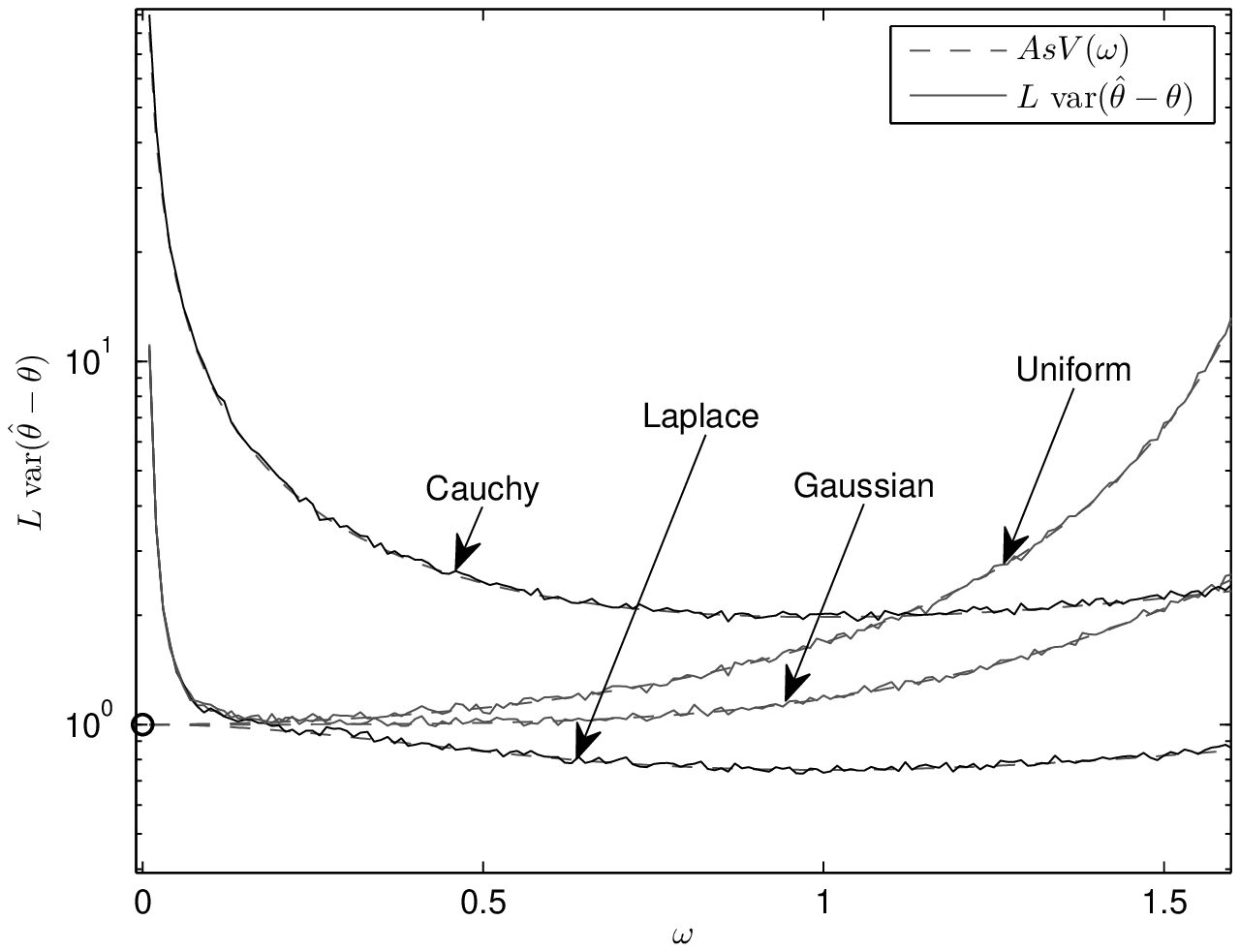}
\caption{Per-sensor Power Constraint, $\sigma_{\eta}^2 = 1$, $\sigma_v^2 = 1$, $L = 500$, $\rho = 1$}\label{fig: asv_w_psc}
\end{center}
\end{minipage}
\end{figure}

\begin{figure}[tb]
\begin{minipage}{1\textwidth}
\centering
\begin{center}
\includegraphics[height=9cm,width=12cm]{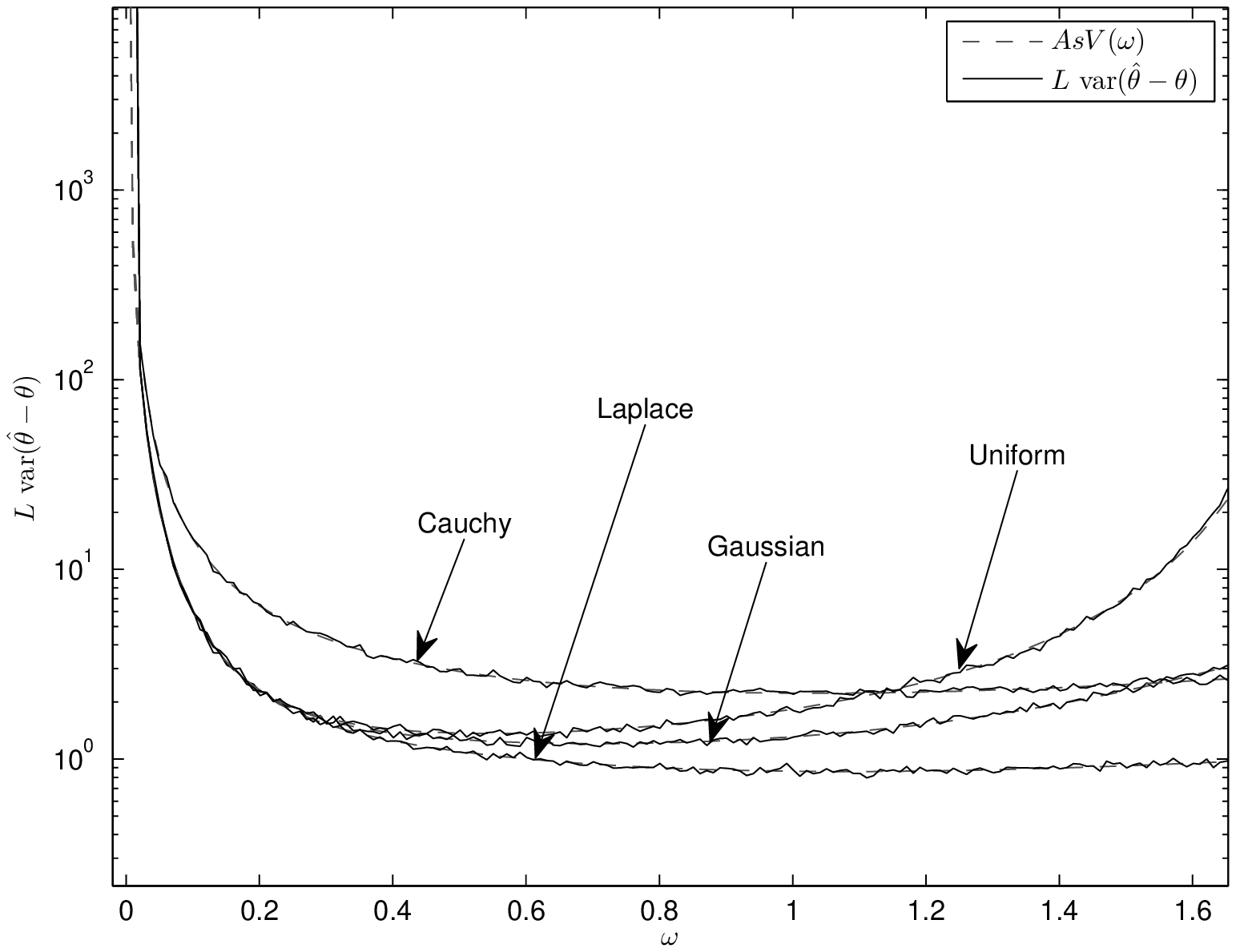}
\caption{Total Power Constraint, $\sigma_{\eta}^2 = 1$, $\sigma_v^2 = 1$, $L = 500$, $P_T = 10$}\label{fig: asv_w_tpc}
\end{center}
\end{minipage}
\end{figure}

\begin{figure}[tb]
\begin{minipage}{1\textwidth}
\centering
\begin{center}
\includegraphics[height=9cm,width=12cm]{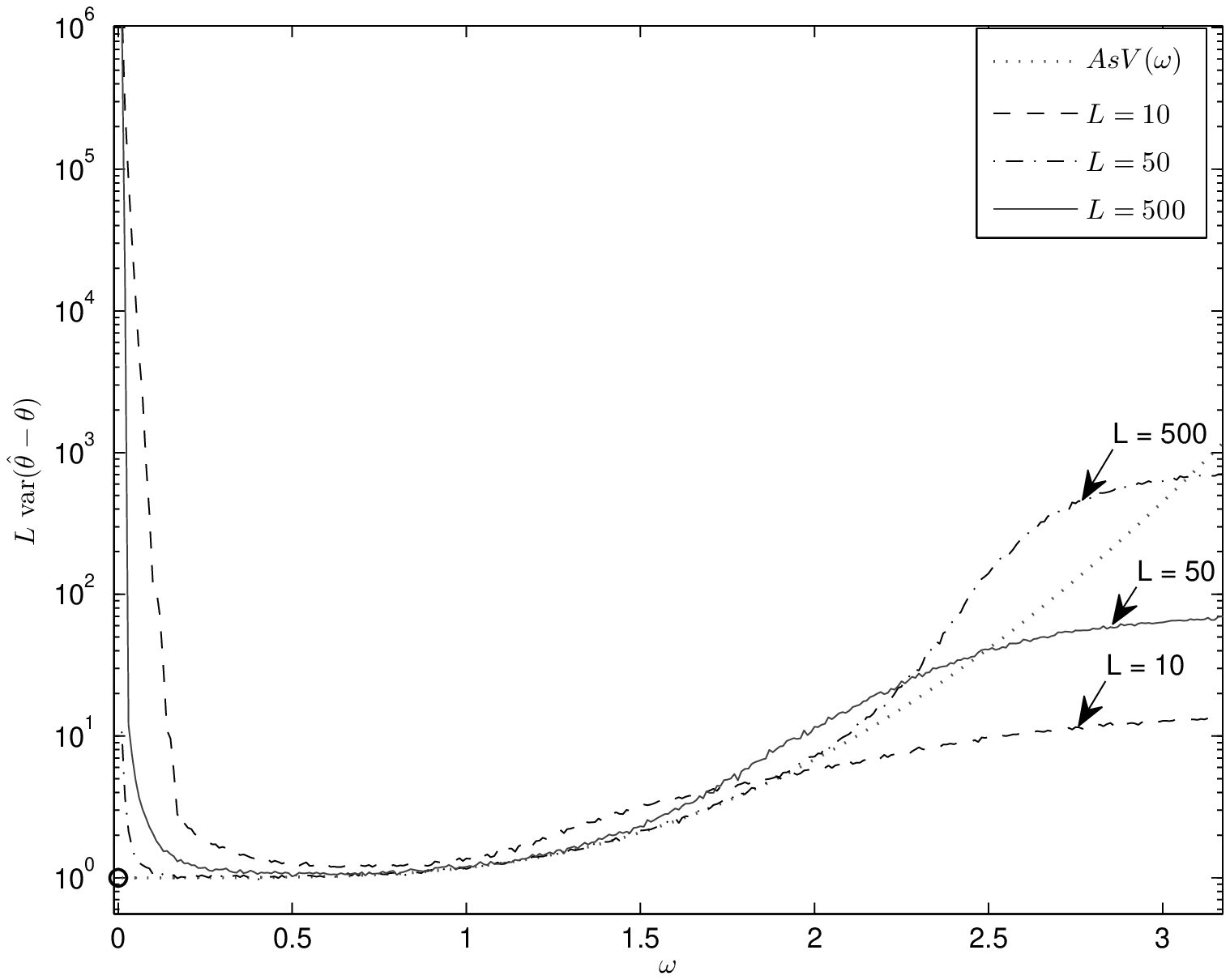}
\caption{Per-sensor Power Constraint, $\sigma_{\eta}^2 = 1$, $\sigma_v^2 = 1$, $\rho = 1$}\label{fig: asv_w_psc_gauss_diffL}
\end{center}
\end{minipage}
\end{figure}

\begin{figure}[tb]
\begin{minipage}{1\textwidth}
\centering
\begin{center}
\includegraphics[height=9cm,width=12cm]{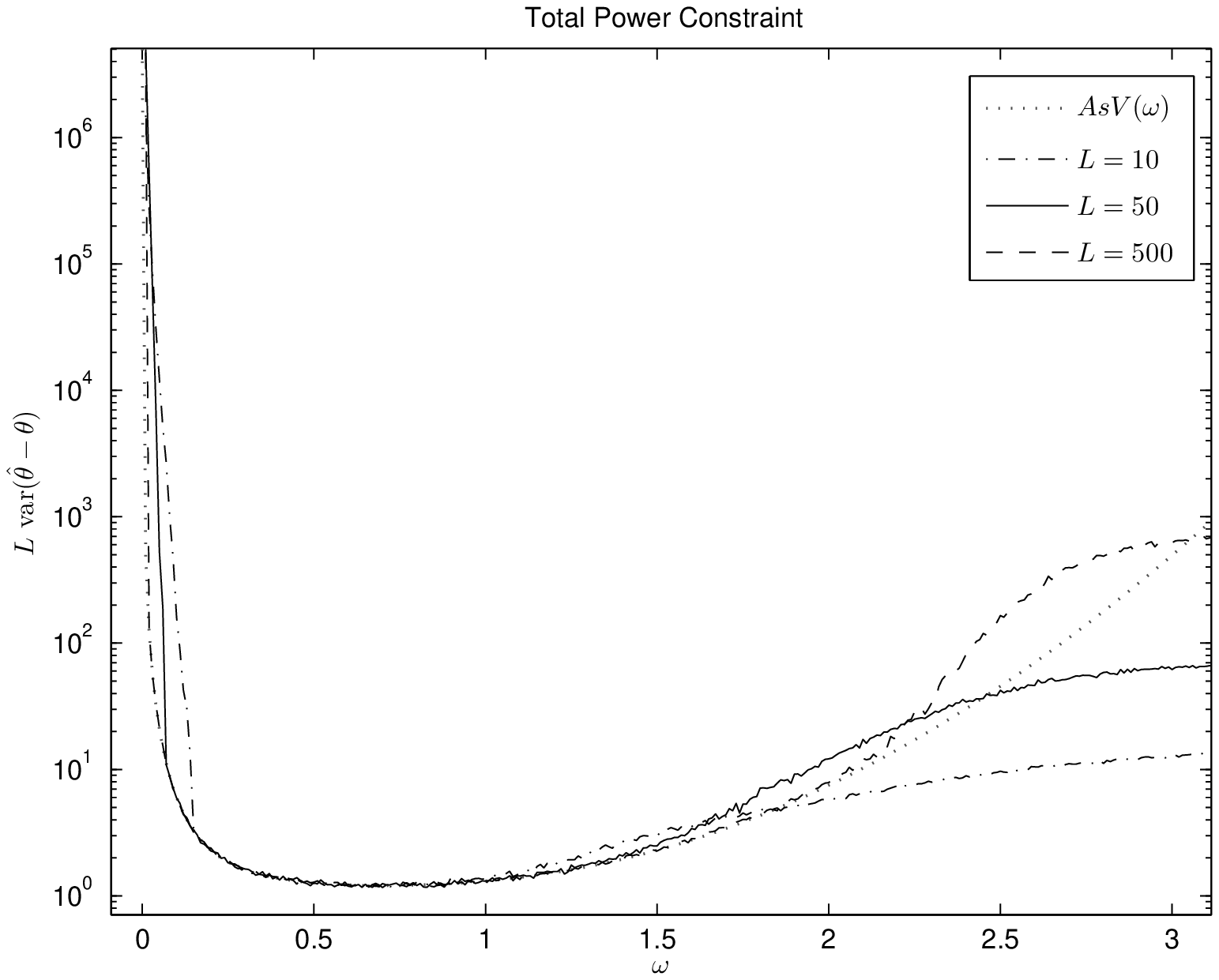}
\caption{Total Power Constraint, $\sigma_{\eta}^2 = 1$, $\sigma_v^2 = 1$, $P_T = 10$}\label{fig: asvw_tpc_gauss_diffL}
\end{center}
\end{minipage}
\end{figure}

In Figures \ref{fig: asv_w_psc} and \ref{fig: asv_w_tpc} we compare $AsV(\w)$ and $L {\rm var}({\hat \theta}-\theta)$ versus $\w$ for
the per-sensor, and total power constraints, respectively.
We begin by acknowledging that the variance of the asymptotic distribution, $AsV(\w)$, and the normalized limiting
variance $L {\rm var}({\hat \theta}-\theta)$ are not always equal in general \cite[pp. 437]{lehmann1998}. However,
as the next two figures show, they are in agreement for the proposed estimator.
The mismatch that occurs for small
$\w$ are due to the number of samples not being sufficiently large for both Figures \ref{fig: asv_w_psc} and \ref{fig: asv_w_tpc}. To focus more on this mismatch, in Figures
\ref{fig: asv_w_psc_gauss_diffL} and \ref{fig: asvw_tpc_gauss_diffL} we consider smaller values of $L$, and
an increased range for $\w$ for the Gaussian sensing noise case. As expected, for reduced values of $L$ the mismatch
increases, especially for small, and large values of $\w$. Note that for the per-sensor power constraint case, although $AsV(\w)$
is bounded near the origin, with finite samples, $L {\rm var}({\hat \theta}-\theta)$ is large for small $\w$,
an effect which is more pronounced for small $L$. This is suggests that for the per-sensor power constraint case $\w$
should not be chosen arbitrarily small, especially when $L$ is small, to avoid this finite-sample artifact.

In Figure \ref{fig: cmpd_gauss_w} we compare $AsV(\w)$ and $L {\rm var}({\hat \theta}-\theta)$ versus $\w$ for
the per-sensor, and total power constraints, respectively, for Middleton Class A noise. In addition to the
agreement of the theory and simulations, these plots illustrate that $AsV(\w)$ need not be a convex, or a
quasiconvex function of $\w$ with a unique local minimum. For all the other noise distributions, $AsV(\w)$
did exhibit a unique local minimum, which was helpful in finding the optimal value of $\w$.

\begin{figure}[tb]
\begin{minipage}{1\textwidth}
\centering
\begin{center}
\includegraphics[height=9cm,width=12cm]{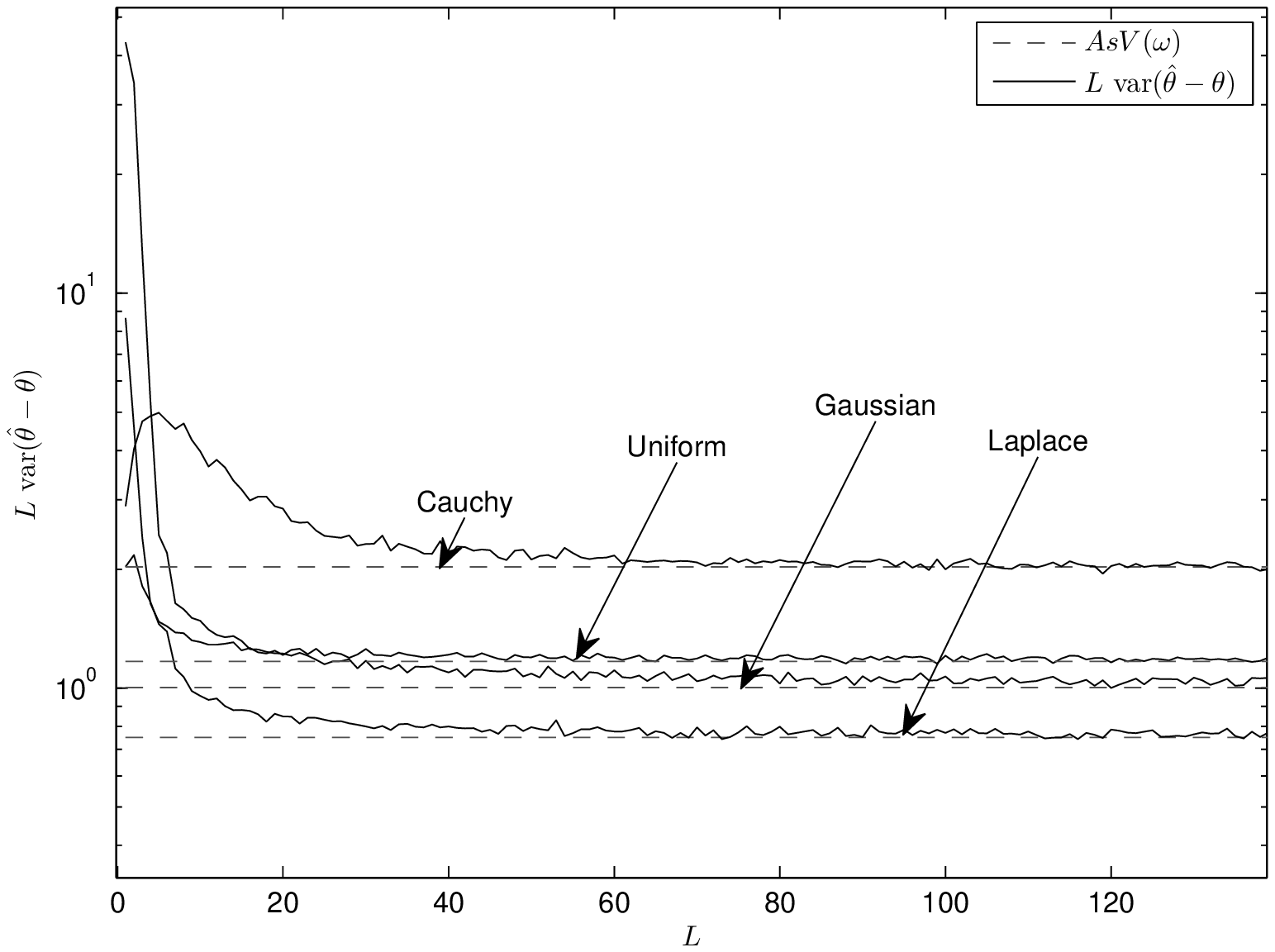}
\caption{Per-sensor Power Constraint, $\sigma_{\eta}^2 = 1$,
$\sigma_v^2 = 1$, $\rho = 1$, $\theta_R = 4$, $\theta =
2$}\label{fig: asv_L_psc}
\end{center}
\end{minipage}
\end{figure}

\begin{figure}[tb]
\begin{minipage}{1\textwidth}
\centering
\begin{center}
\includegraphics[height=9cm,width=12cm]{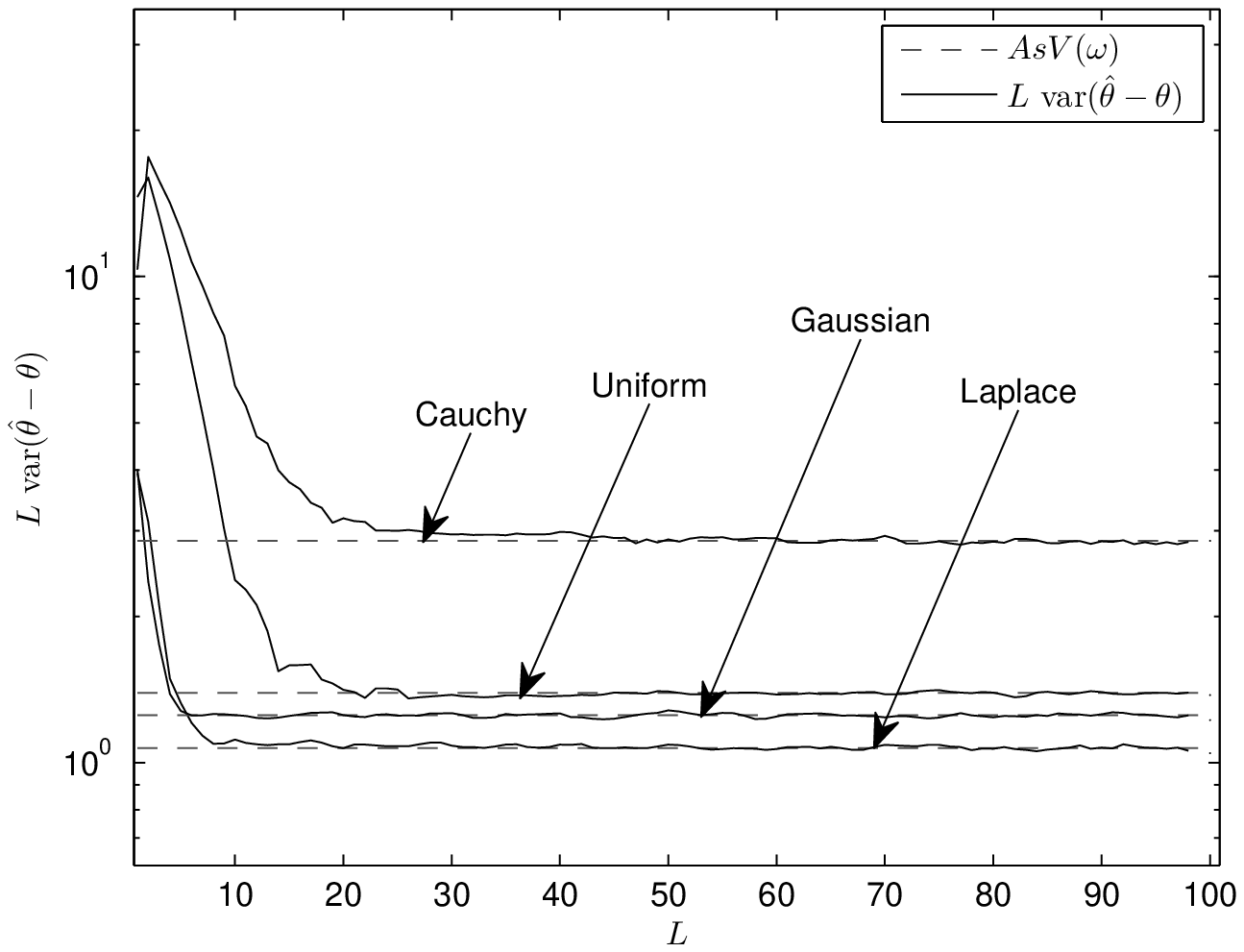}
\caption{Total Power Constraint, $\sigma_{\eta}^2 = 1$, $\sigma_v^2 = 1$, $P_T = 10$, $\theta_R = 12$, $\theta = 2$}\label{fig: asv_L_tpc}
\end{center}
\end{minipage}
\end{figure}

\begin{figure}[tb]
\begin{minipage}{1\textwidth}
\centering
\begin{center}
\includegraphics[height=9cm,width=12cm]{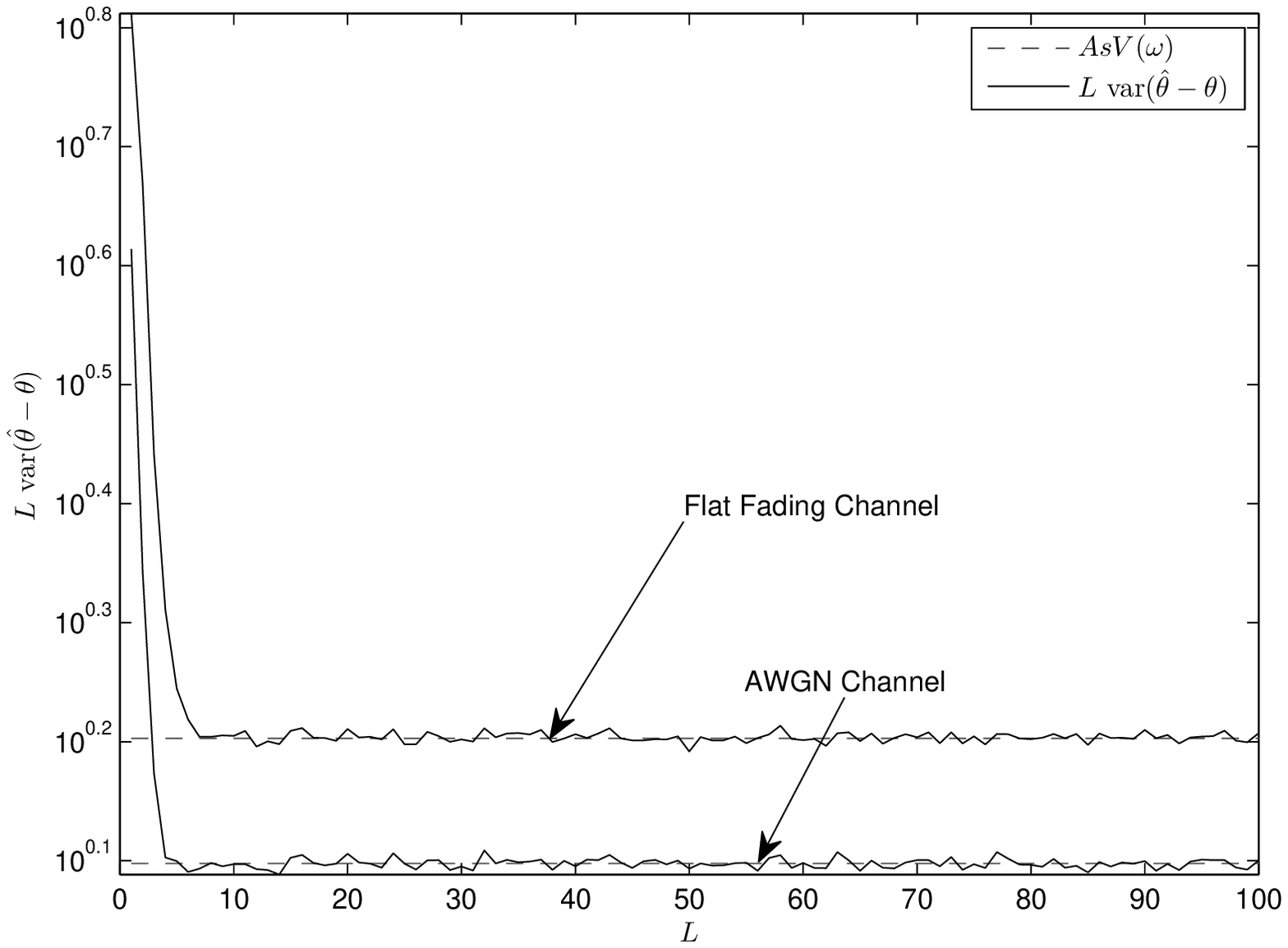}
\caption{Total Power Constraint, E$[|h|^2] = 1$, $\sigma_{\eta}^2 = 1$, $\sigma_v^2 = 1$, $P_T = 10$, $\theta_R = 12$, $\theta = 2$, }\label{fig: fading}
\end{center}
\end{minipage}
\end{figure}

Figures \ref{fig: asv_L_psc} and \ref{fig: asv_L_tpc} show $L {\rm var}({\hat \theta}-\theta)$ versus $L$ for
the per-sensor, and total power constraints, respectively.
The optimal value of $\w$ that minimizes the $AsV$ is chosen for the total power constraint case.
For the per-sensor
power constraint case in Figure \ref{fig: asv_L_psc}, we did not use the minimizer of $AsV(\w)$ 
due to the  aforementioned finite-sample effects. Instead,
the value of $\w$ is chosen to minimize $L {\rm var}({\hat \theta}-\theta)$
in Figure \ref{fig: asv_w_psc} (which
assumes $L=500$) and applied to all values of $L$ in Figure \ref{fig: asv_L_psc}.
It is seen that convergence occurs slower for the heavy-tailed Cauchy distribution.
At about $L=50$, all cases converge for both Figures.
Figure \ref{fig: fading} illustrates the effect of Rayleigh fading on the performance for Gaussian sensing and channel noise.
It is seen that $L {\rm var}({\hat \theta}-\theta)$ converges to their theoretically predicted asymptotic value with a ratio
of about $4/\pi$ compared to the non-fading case.

\begin{figure}[tb]
\begin{minipage}{1\textwidth}
\centering
\begin{center}
\includegraphics[height=9cm,width=12cm]{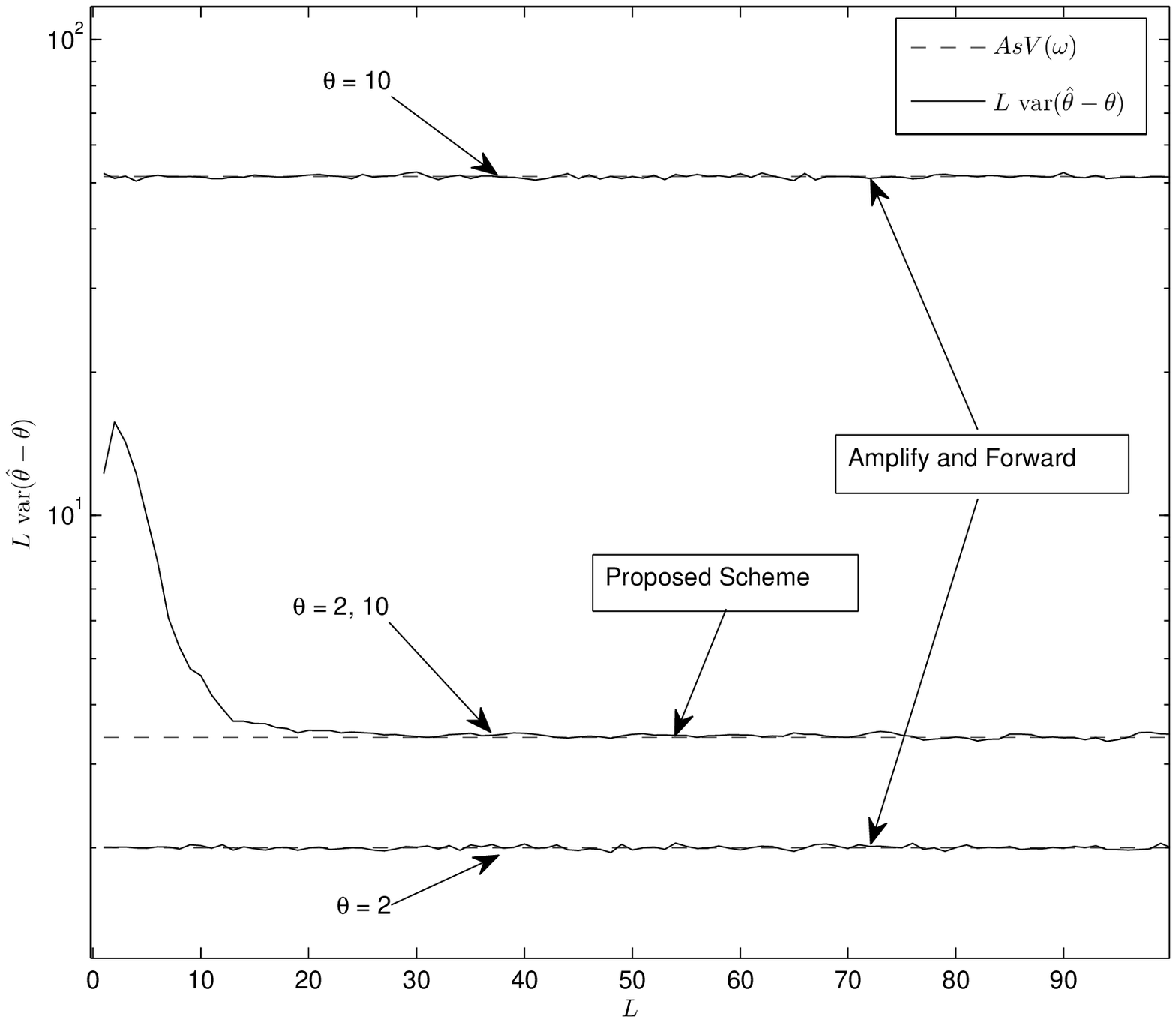}
\caption{Total Power Constraint, $\sigma_{\eta}^2 = 1$, $\sigma_v^2 = 1$, $P_T = 10$, $\theta_R = 12$}\label{fig: anf_theta_comp}
\end{center}
\end{minipage}
\end{figure}

In Figure \ref{fig: anf_theta_comp} the proposed scheme is compared with AF. The performance of the proposed
approach is seen to be both better and worse than AF depending on the value of $\theta$.
Another interesting aspect of Figure \ref{fig: anf_theta_comp} is the flatness of the curves for the AF
case. This can be seen by finding the variance of equation (\ref{asvaf}), which is a constant function of $L$.
In contrast, the normalized variance for the proposed estimator is seen to depend on $L$ in Figure \ref{fig: anf_theta_comp}.

\begin{figure}[tb]
\begin{minipage}{1\textwidth}
\centering
\begin{center}
\includegraphics[height=9cm,width=12cm]{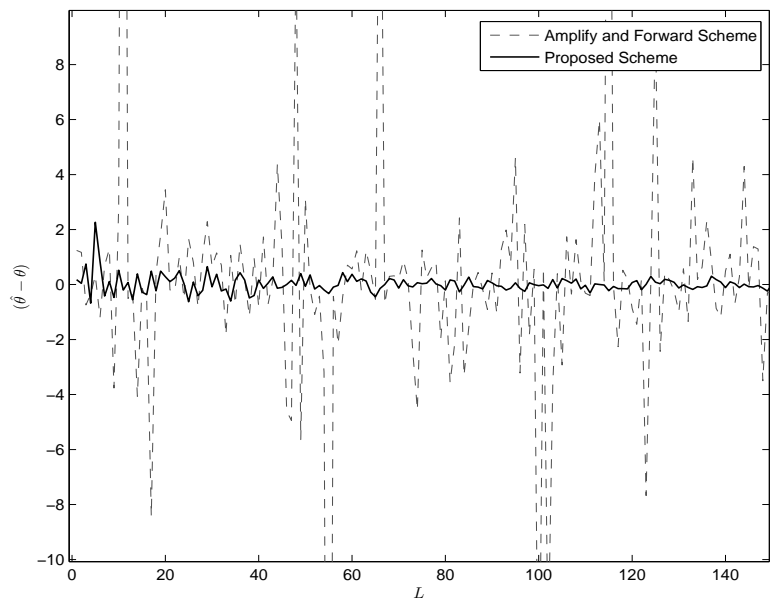}
\caption{Total Power Constraint with Cauchy distributed sensing noise, $\sigma_{\eta}^2 = 1$, $\sigma_v^2 = 1$, $P_T = 10$, $\theta_R = 12$, $\theta = 2$}\label{fig: anf_cauchy2}
\end{center}
\end{minipage}
\end{figure}

To illustrate the robustness of the proposed estimator Figure \ref{fig: anf_cauchy2} compares
it with AF for Cauchy sensing noise. One realization of the estimation error is plotted for each value of $L$ to illustrate that in the presence of Cauchy noise,
the performance of AF does not converge despite the increase in $L$, whereas the proposed estimator is consistent.

\section{Conclusions} \label{Conclusions}
A distributed estimation scheme relying on constant modulus transmissions from the sensors is proposed over
Gaussian multiple access channels. The instantaneous transmit power does not depend on the random sensing
noise, which is a desirable feature for low-power sensors with limited peak power capabilities. In the i.i.d.
case, the estimator is shown to be strongly consistent for any sensing or channel noise distribution. In the
non-identically distributed case, a bound on the variances is shown to be a sufficient condition for
strong consistency. The asymptotic variance is derived, and shown to depend on the characteristic function
of the sensing noise which is bounded for the general case, and also optimized with respect to $\w$ for
various noise distributions. In addition to the desirable constant-power feature, the proposed estimator
is robust to impulsive noise, and remains consistent even when the mean and variance of the sensing noise
does not exist.
It is argued that over Gaussian multiple access channels, the AF estimator is effectively a noisy sample
mean of the sensed data. For sensing noise distributions for which the sample mean is highly suboptimal
or inconsistent, the proposed estimator is shown to outperform AF.
The effect of fading is also considered, and shown to effect the asymptotic variance by a constant
fading penalty factor.

\section*{Appendix 1: Proof of Theorem \ref{thm2}} \label{sec: Appendix1}

We begin by observing that the $2 \times 1$ vector sequence, $\sqrt{L}\left[\nrmfrxR-\avgnrmfrx^R \hspace{0.1 in} \nrmfrxI-\avgnrmfrx^I\right]$ is asymptotically normal with zero mean, due to the central limit theorem. The elements of its asymptotic covariance matrix can be calculated to be,
\begin{align} \label{eqn: covars}
\Sigma_{11} := \powcnst\left[ \cos^2(\w \ttheta)\cdot v_c+ \sin^2(\w \ttheta)\cdot v_s\right]+\frac{\sigv}{2} \\ \Sigma_{22} := \powcnst\left[\cos^2(\w \ttheta)\cdot v_s+\sin^2(\w \ttheta) \cdot v_c \right]+\frac{\sigv}{2} \\ \Sigma_{12} = \Sigma_{21}:= \powcnst \sin(\w \ttheta) \cos (\w \ttheta)(v_c-v_s) \end{align} where, for brevity we have $v_c:= {\rm{var}}(\cos \w \ieta)= (1/2)+ \Chardeta/2-\Charseta$ and $v_s:= {\rm{var}}(\sin \w \ieta) = (1/2)-\Chardeta/2$. Applying  \cite[Thm 3.16]{porat1994} the asymptotic variance is given by
\begin{equation}\label{eqn: asymp_var2}
AsV = G_1^2 \Sigma_{11}+2 G_1G_2\Sigma_{12}+G_2^2 \Sigma_{22}
\end{equation}where
\begin{align}
G_1&:= \frac{\partial \frac{1}{\w}\tan^{-1}\frac{x}{y}}{\partial x}\bigg|_{{x}=\avgnrmfrx^R} = -\frac{1}{\w} \frac{1}{1+\tan^2(\w \ttheta)}\frac{\tan(\w \ttheta)}{\sqrt{\powcnst} \Chareta \cos(\w \ttheta)}\\
G_2 &:= \frac{\partial \frac{1}{\w}\tan^{-1}\left( \frac{x}{y}\right)}{\partial y} \bigg|_{y = \avgnrmfrx^I} = \frac{-G_1}{\tan{\w \ttheta}}
\end{align} Substituting in (\ref{eqn: asymp_var2}) and simplifying we obtain the theorem.

\section*{Appendix 2: Proof of Theorem \ref{thm4}} \label{sec: Appendix2}
Using the bound, $\varphi_{\ieta}(\w) \geq 1- \sigma_{\ieta}^2 \w^2/2$,
we have for all $\w$, $\Chardeta \geq 1-2\sigeta \w^2$, and for $\w<\sqrt{2}/\sigma_{\eta}$, $\Charseta \geq \left(1-\sigeta \w^2/2\right)^2$.
Substituting in (\ref{eqn: Asymp-var}) we have for $\w<\sqrt{2}/\sigma_{\eta}$,
\begin{equation}\label{eqn: ineq3}
AsV(\w) \leq \frac{\left[\frac{\sigeta}{\powcnst}+2\sigeta\w^2 \right]}{2\w^2\left(1-\frac{\sigeta \w^2}{2} \right)^2} \;.
\end{equation}
Recall that $2\pi/\thetarng< \sqrt{2}/\sigma_{\eta}$ by assumption. Therefore, upper bound (\ref{eqn: ineq3})
is valid over the entire range of $\w$ values which involves the minimization in (\ref{eqn: opt_asymp_var}).
We can therefore minimize both sides of (\ref{eqn: ineq3}) over $\w \in (0, 2\pi/\thetarng]$. Substituting for convenience $\beta \leftarrow \w^2 \sigeta$ we have
\begin{equation}\label{eqn: ineq4}
AsV^{\ast} \leq \min_{\beta \in (0, (2\pi\sigma_{\eta}/\thetarng)^2]} \frac{\left[\frac{\sigv}{\powcnst}+2\beta \right]}{2\beta(1-\beta/2)^2}\sigeta.
\end{equation}
The unconstrained minimum can be found by differentiating (\ref{eqn: ineq4}) and is given by $c/8$, with a corresponding minimum given by the right hand side (rhs) of (\ref{eqn: ineq1}).
It can be checked that $c/8$ is the unique minimum of the unconstrained problem. This shows that if $(2\pi\sigma_{\eta}/\thetarng)^2> c/8$ then the rhs of (\ref{eqn: ineq4}) is given by the rhs of (\ref{eqn: ineq1}).

To show (\ref{eqn: ineq2}), recall that $c/8$ is the unique unconstrained minimum of the objective on the
rhs of (\ref{eqn: ineq4}). This implies that as a function of $\beta$ it is non-increasing
over $(0,c/8)$ so that when $(2\pi\sigma_{\eta}/\thetarng)^2 < c/8$ the minimum over
$[0, (2\pi\sigma_{\eta}/\thetarng)^2]$ is achieved at $\beta = (2\pi \sigma_{\eta}/\thetarng)^2$
which is the rhs of (\ref{eqn: ineq2}).
\bibliographystyle{IEEEtran}
\bibliography{robustest}
\newpage

\end{document}